\documentclass[aps,pre,preprintnumbers,showpacs,superscriptaddress]{revtex4}

\usepackage{amsmath}
\usepackage{graphicx}

\begin{document}

\title{Network Coordination and Synchronization in a Noisy Environment with Time Delays}

\author{D. Hunt}
\affiliation{Department of Physics, Applied Physics, and Astronomy}
\affiliation{Social and Cognitive Networks Academic Research Center}

\author{B.K. Szymanski}
\affiliation{Social and Cognitive Networks Academic Research Center}
\affiliation{Department of Computer Science \\
Rensselaer Polytechnic Institute, 110 8$^{th}$ Street, Troy, NY 12180--3590, USA}

\author{G. Korniss\footnote{Corresponding author. korniss@rpi.edu}}
\affiliation{Department of Physics, Applied Physics, and Astronomy}
\affiliation{Social and Cognitive Networks Academic Research Center}


\begin{abstract}
We study the effects of nonzero time delays in stochastic
synchronization problems with linear couplings in complex networks.
We consider two types of time delays: transmission delays between
interacting nodes and local delays at each node (due to processing,
cognitive, or execution delays). By investigating the underlying
fluctuations for several delay schemes, we obtain the
synchronizability threshold (phase boundary) and the scaling
behavior of the width of the synchronization landscape, in some
cases for arbitrary networks and in others for specific weighted
networks. Numerical computations allow the behavior of these
networks to be explored when direct analytical results are not
available. We comment on the implications of these findings for
simple locally or globally weighted network couplings and possible
trade-offs present in such systems.
\end{abstract}


\date{\today}
\maketitle

\section{Introduction}

Since the classic works by Kalecki \cite{Kalecki_1935} and Frisch \&
Holme \cite{Frisch_1935} on the emergence of macro-economical
patterns (business and economics cycles), it has been
well-established that time delays occurring on microscopic scales
can have profound effects on the global response of complex systems.
Among other early key results were the works by Hutchinson
\cite{Hutchinson_1948} and May \cite{May_1973}, showing that time
delays can have fundamental impact on logistic growth in population
dynamics \cite{Ruan_2006}. The importance of time delay becomes even
more explicit in interacting individual- or agent-based models
\cite{Saber_IEEE2007,Sipahi_IEEE2011,Jadbabaie_IEEE2010,Huberman_IEEE1991,Strogatz_PRE2003},
where the delays can correspond to time scales in the interactions (e.g. transmission
delays) or to time scales of the local decision and execution by the
individuals. In this paper, we consider the simplest -- yet
fundamental -- model for such networked systems, taking into
consideration the effects of the network topology and couplings,
noise, and time delays \cite{HuntPRL}. This paper provides an
extended account of our recent Letter \cite{HuntPRL}, providing more
details, generalizations and comparisons to certain weighted
networks, and considering different types of delays.

In network synchronization \cite{Arenas_PhysRep2008}, coordination,
or consensus problems \cite{Saber_IEEE2007}, individuals or entities
represented by nodes in the network attempt to adjust their local
state variables (e.g., pace, load, phase, or orientation) in a
decentralized fashion. (In this paper, we use the terms
synchronization, coordination, and consensus synonymously in this
broader sense.) Nodes interact or communicate only with their local
neighbors in the network, often with the intention to improve global
performance. These couplings can be represented by directed or
undirected, weighted or unweighted links. Applications of the
corresponding models range from physics, biology, computer science
to control theory, including synchronization problems in distributed
computing \cite{GK_Science2003}, symbolic dynamics
\cite{Jalan_PLA2010}, congestion control in communication networks
\cite{GK_PRE2007,Korniss_Springer2012,Johari_IEEE2001,Saber_IEEE2004,Saber_IEEE2007}
and in vehicular traffic \cite{Orosz_PRSA2006,Orosz_PTRSA2010},
flocking animals
\cite{Reynolds_CG1987,Vicsek_PRL1995,Cucker_IEEE2007}, bursting
neurons \cite{Izhikevich_SIAM2001}, and cooperative control of
vehicle formation \cite{Fax_IEEE2004}.

Synchronization, coordination, or consensus in complex networks cuts
across numerous fields that address global behavior through
decentralized local actions facilitated by sparse interactions.
There has already been much investigation into the efficiency and
optimization of synchronization
\cite{Arenas_PhysRep2008,Barahona_PRL2002,Nishikawa_PRL2002,LaRocca_PRE2008,LaRocca_PRE2009}
in weighted \cite{Zhou_PRL2006,GK_PRE2007,Korniss_Springer2012} and directed
\cite{Saber_IEEE2007,Nishikawa_PRE2006,Nishikawa_2009} topologies.
Because of limitations in communication, transportation, processing,
or cognitive resources, the local information on the state of the
network neighborhood may not always be current, nor is it even given
for the same instant at a past time for all components. These time
delays can have drastic effects on system behavior
\cite{Sipahi_IEEE2011} and further complicate predictability of the
network's global performance.

The impact of time delays on stochastic differential equations
involving a single stochastic variable, with recent applications to
postural sway \cite{Milton_EPL2008,Milton_PTRSA2009}, stick
balancing at a fingertip \cite{Cabrera_PRL2002,Cabrera_CMP2006}, and
the scaling of congestion window in internet protocols
\cite{Ott_2006}, have been investigated in the past two decades
\cite{Kuechler_SSR1992,Ohira_PRE2000,Frank_PRE2001,Pikovsky_PRL2001}.
Here, we focus on the interplay of network topology, couplings,
noise, and time delays. Our motivation is to understand how
network-connected individuals contribute to global goals by
performing delayed actions and/or using delayed information
facilitated by local interactions in a noisy environment.

The phenomena of spontaneous synchronization, coordination, or
consensus arise in a variety of disciplines
\cite{Korniss_Springer2012,Arenas_PhysRep2008,Saber_IEEE2007,Sipahi_IEEE2011}.
For example, it describes the consensus that arises in bird flocks
as each bird makes velocity adjustments to match the group, which is
crucial in accomplishing such tasks as avoiding predators
\cite{Cucker_IEEE2007, Vicsek_PRL1995}. Similarly, it can be applied
to a collection of autonomous vehicles working cooperatively to
carry out a task \cite{Saber_IEEE2004}. Risk can be managed without
central governance in uncertain environments through the
synchronicity or spontaneous cooperation of individuals. This
appears in economics when considering stock trades
\cite{Saavedra_PNAS2012}; in ecology there is the reward of
reproduction and the danger of predation for chirping cicadas and
flashing fireflies \cite{fireflies}. While there are adversarial
relationships between individual participants, there are still
mutual benefits (predictive insight or bodily protection) from the
collective behavior. Massively parallel and distributed computing
schemes require synchronization across processors
\cite{GK_Science2003,Korniss_PRL2000,Guclu_PRE2006,Guclu_Chaos2007} in order to
avoid diverging progressions of simulation time but must be balanced
with the cost of communication. Synchronization of coupled phase
oscillators \cite{PhysRevE.65.026139} (the Kuramoto model
\cite{Kuramoto}) has many applications, recently to spatial patterns
in flashing microfluidic arrays \cite{PhysRevE.83.046206} and to
circuits comprised of optomechanical arrays \cite{optomechanical}.
In neural networks, time delays critically affect the
synchronization of excitatory fronts
\cite{Chen_EPL2008,Chen_PRE2009,Chen_PLOS2011}. All these examples
are instances of a group coming to consensus \cite{Saber_IEEE2007}
without an omniscient global operator. They fundamentally rely on
the communication between individuals, which may be (and often is)
sent through noisy channels
\cite{Ohira_PRE2000,Frank_PRE2001,Pikovsky_PRL2001}.

\subsection{The Model}

In the model we consider here, the state of each node $i$ is
described by a local scalar state variable $h_i$. In stochastic
network coordination/consensus problems, nodes locally adjust their
state in an attempt to match that of their neighbors through linear
couplings in the presence of noise. However, they react to the
information or signal received from their neighbors with some time
lag, and the evolution of the states of the nodes is governed by the
differential time-delay equations
\begin{equation}
\partial_t h_i(t) = -\sum_j C_{ij}[h_i(t - \tau^{\rm o}_{i}) - h_j(t - \tau^{\rm o}_{i} - \tau^{\rm tr}_{ij})] + \eta_i(t) \;.
\label{diffEqGeneral}
\end{equation}
Here, $C_{ij}$ is the coupling strength between nodes $i$ and $j$,
and $\eta_i$ is the noise present at node $i$, satisfying
$\langle\eta_i(t)\eta_j(t')\rangle = 2D\delta_{ij}\delta(t - t')$,
where $D$ is the noise intensity. In general, the time delays can be
heterogeneous, depending on the properties and network locations of
both nodes: $\tau^{\rm o}_{i}$ is the local delay at node $i$,
corresponding to processing, cognitive, or execution delays, while
$\tau^{\rm tr}_{ij}$ is the transmission delay between nodes $i$ and
$j$. Without the noise term, the above equation is often referred to
as the (deterministic) consensus problem
\cite{Saber_IEEE2004,Saber_IEEE2007} on the respective network. In
this sense, the networked agents try to coordinate or reach an
agreement or balance regarding a certain quantity of interest.

A standard measure of synchronization, coordination, or consensus in a noisy environment is the width \cite{GK_PRE2007,GK_Science2003}
\begin{equation}
\langle w^2(t) \rangle = \biggl\langle \frac{1}{N}\sum_{i = 1}^N[h_i(t) - \bar h(t)]^2\biggr\rangle \;,
\label{w2_def}
\end{equation}
where $\bar{h}(t)=(1/N)\sum_{i=1}^{N}h_i(t)$ is the global average of the local state variables and $\langle\ldots\rangle$ denotes an ensemble average over the noise.
A network is ``synchronizable'' if it asymptotically reaches a steady state with a finite width, i.e. $\langle w(\infty)\rangle < \infty$.
When the network is well synchronized (or coordinated), the values $h_i$ for all nodes are near the global mean $\bar h$ and the width is small.

\subsection{Coordination without Time Delays}

Without time delays, Eq.~(\ref{diffEqGeneral}) takes the form \begin{equation}\partial_t h_i(t) = -\sum_j C_{ij}[h_i(t) - h_j(t)] + \eta_i(t) = -\sum_j \Gamma_{ij} h_j(t) + \eta_i(t)
\label{diffEqNoDelay}
\end{equation}
where $\Gamma_{ij} = \delta_{ij}\sum_{l} C_{il} - C_{ij}$ is the
network Laplacian. Eq.~(\ref{diffEqNoDelay}) is a multivariate
Ornstein-Uhlenbeck process \cite{Gardiner_1985} and is also referred
to as the Edwards-Wilkinson process \cite{EW} on a network
\cite{GK_Science2003,GK_PRE2007}. Starting from a flat initial
profile $\{h_i(0)=0\}_{i=1}^{N}$ for symmetric couplings, one can
show that the width evolves as \cite{Gardiner_1985}
\begin{equation}
\langle w^2(t)\rangle = \frac{D}{N}\sum_{k = 1}^{N - 1}\frac{(1 - e^{-2\lambda_k t})}{\lambda_k} \;,
\label{widthNoDelay}
\end{equation}
where $\lambda_k$, $k=0,1,2,\ldots,N-1$, are the eigenvalues of the
network Laplacian. Note that as a result of measuring the local
state variables $h_i$ from the mean $\bar{h}$ in Eq.~(\ref{w2_def}),
the singular contribution of $\lambda_{0} = 0$ (associated with the
uniform mode) automatically cancels out from the sum in
Eq.~(\ref{widthNoDelay}). Thus, a {\em finite connected} network is
always synchronizable with steady-state width
\begin{equation}
\langle w^2(\infty)\rangle = \frac{D}{N}\sum_{k = 1}^{N - 1}\frac{1}{\lambda_k} \;.
\label{w2_steady}
\end{equation}
In the limit of infinite network size, however, network ensembles with a vanishing (Laplacian) spectral gap may become unsynchronizable, depending on the details of the small-$\lambda$ behavior of the density of eigenvalues \cite{Arenas_PhysRep2008,GK_Science2003,GK_PRE2007}.
This type of singularity is common in purely spatial networks (in particular, in low dimensions) where the relevant response functions and fluctuations diverge in the long-wavelength (small-$\lambda$) limit \cite{GK_Science2003,Goldenfeld_1992}.
In complex networks \cite{Barab_sci,Watts_Nature1998,BarabREV,MendesREV} these singularities are typically suppressed as a result of sufficient amount of randomness in the connectivity pattern generating a gap or ``pseudo" gap. \cite{Monasson_EPJB1999,GK_Science2003,Barahona_PRL2002,Kozma_UGA2004,Kozma_PRL2004,Kim_PRL2007}.

As is also clear from Eq.~(\ref{w2_steady}), synchronization or
coordination can be arbitrarily improved in this case of no time
delays, e.g., by uniformly increasing the coupling strength by a
factor of $\sigma > 1$, resulting in $C_{ij} \rightarrow \sigma
C_{ij}$ ($\lambda_k \rightarrow \sigma\lambda_k$) and yielding
\begin{equation}
\langle w^2(\infty)\rangle_{\sigma} = \frac{1}{\sigma} \langle w^2(\infty)\rangle_{\sigma=1} \;.
\end{equation}
The stronger the effective coupling $\sigma$ (e.g., achieved by more frequent communications in real networks), the better the synchronization; the width is a monotonically decreasing function of $\sigma$.

\section{Uniform Local Time Delays}
\label{section_uniformDelays}

We first consider the case with symmetric coupling $C_{ij} = C_{ji}$ when transmission delays are negligible ($\tau^{\rm tr}_{ij} = 0$) and local delays are uniform ($\tau^{o}_{i} \equiv \tau$).
Then Eq.~(\ref{diffEqGeneral}) is governed by a single uniform time delay \cite{HuntPRL}
\begin{equation}
\partial_t h_i(t) = -\sum_{j = 1}^N C_{ij}[h_i(t - \tau) - h_j(t - \tau)] + \eta_i(t)
= -\sum_{j = 1}^N \Gamma_{ij}h_j(t - \tau) + \eta_i(t) \; .
\label{diffEqUniform}
\end{equation}
This equation has a similar form to that of
Eq.~(\ref{diffEqNoDelay}) but with the inclusion a delay $\tau$.

\subsection{Eigenmode Decomposition and Scaling}

By diagonalizing the symmetric network Laplacian $\Gamma$, the above set of equations of motion decouples into separate modes
\begin{equation}
\partial_t\tilde h_k(t) = -\lambda_k\tilde h_k(t - \tau) + \tilde\eta_k(t) \;,
\label{hk_evol}
\end{equation}
where $\lambda_k$ ($k=0,1,2,\ldots,N-1$) are the eigenvalues of the network Laplacian, and $\tilde{h}_k$ and $\tilde{\eta}_k$ are the time-dependent components of the state and noise vectors, respectively, along the $k$-th eigenvector.
Thus, the amplitude $\tilde{h}_k$ of each mode (with the exception of the uniform mode with $\lambda_0 = 0$) is governed by the same type of stochastic delay-differential equation
\begin{equation}
\partial_t\tilde{h}(t) = -\lambda \tilde{h}(t - \tau) + \tilde{\eta}(t) \;,
\label{h_evol}
\end{equation}
with $\lambda > 0$, where we temporarily drop the index $k$ of the specific eigenmode for transparency and to streamline notation.

While the above stochastic delay-differential equation has an exact
stationary solution for the stationary-state variance
\cite{Kuechler_SSR1992,Frank_PRE2001}, we first review the formal
solution \cite{Amann_PhysA2007,HuntPRL} which provides some insights
and connections between the solutions of the underlying
characteristic equation and the existence (and the scaling) of the
stationary-state fluctuations of the stochastic problem. The formal
solution can also be applied to more general linear (or linearized)
coordination problems with multiple time delays \cite{HuntPLA}, and
can serve as the starting point to extract the asymptotic behavior
\cite{Hod} near the singular points (synchronization boundary).

\begin{figure}[t]
\vspace{1.0truecm}
\centering
\includegraphics[scale=0.4]{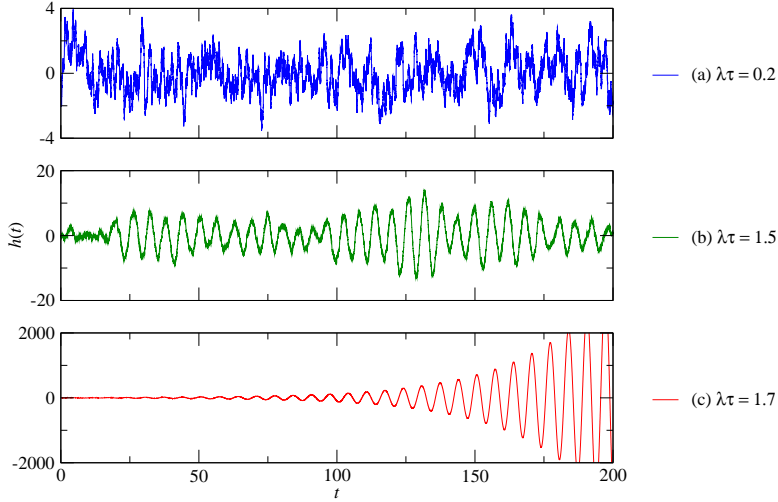}
\caption{(Color online) Time evolution of an individual mode obtained by
numerically integrating Eq.~(\ref{h_evol}) with $\lambda$$=$$1$, $D$$=$$1$, and $\Delta t$$=$$0.001$ for
several delays chosen to show the various behaviors across the
separating/critical points $\lambda\tau$$=$$1/e$ and $\pi/2$;
(a) $\lambda\tau$$=$$0.2<1/e$,
(b) $1/e < \lambda\tau$$=$$1.5<\pi/2$, and
(c) $\lambda\tau$$=$$1.7>\pi/2$.}
\label{uniformModeEvol}
\end{figure}

\begin{figure}[t]
\vspace{1.0truecm}
\centering
\includegraphics[scale=0.5]{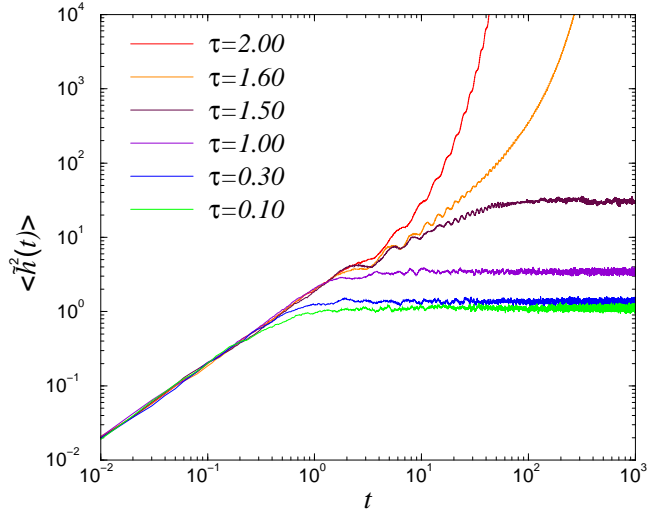}
\caption{(Color online) Time series of the fluctuations of a single mode ($\lambda$$=$$1$) averaged over $10^4$ realizations of noise
(with $D$$=$$1$) by numerically integrating Eq.~(\ref{h_evol}) with $\Delta t$$=$$0.01$ for different delays
(from bottom to top in increasing order of $\tau$).}
\label{uniformModeFluctEvol}
\end{figure}

Performing standard Laplace transform on Eq.~(\ref{h_evol}) [with
$\hat{h}(s)=\int_{0}^{\infty}e^{-st}\tilde{h}(t)dt$], the
characteristic equation associated with its homogeneous
(deterministic) part becomes
\begin{equation}
g(s) \equiv s + \lambda e^{-s\tau} = 0 \;.
\label{char_eq_uniform}
\end{equation}
As shown in Appendix~\ref{appendix_Fluctuations}, (with $h(t) \equiv 0$ for $t \leq 0$) the time-dependent fluctuations can be written formally as
\begin{equation}
\langle \tilde{h}^2(t)\rangle =
\sum_{\alpha,\beta} \frac{-2D( 1- e^{(s_{\alpha} + s_{\beta})t} )}{g^{'}(s_{\alpha}) g^{'}(s_{\beta})(s_{\alpha} + s_{\beta})} \;.
\end{equation}
Hence, they remain {\em finite} (i.e., a stationary distribution exists) if
\begin{equation}
{\rm Re}(s_\alpha) < 0 \;,
\label{Re_s}
\end{equation}
for {\em all} $\alpha$, where $s_\alpha$, $\alpha=1,2,\ldots$, are the solutions of the characteristic equation, Eq.~(\ref{char_eq_uniform}), on the complex plane.
We can explicitly make the simplification
\begin{equation}
\langle \tilde{h}^2(\infty)\rangle  =
  \sum_{\alpha,\beta} \frac{-2D}{g^{'}(s_{\alpha}) g^{'}(s_{\beta})(s_{\alpha} + s_{\beta})}
= \sum_{\alpha,\beta} \frac{-2D}{(1 + \tau s_{\alpha})(1 + \tau s_{\beta})(s_{\alpha} + s_{\beta})} \;.
\label{h2_st}
\end{equation}
Eq.~(\ref{char_eq_uniform}) is perhaps the oldest and most
well-known (transcendental) characteristic equation from the theory
of delay-differential equations
\cite{Saber_IEEE2004,Frisch_1935,Hayes_1950,Ruan_2006}, with the
linear stability analysis of numerous nonlinear systems reducing to
this one. It has an infinite number of (in general, complex)
solutions for $\tau > 0$ and the condition in Eq.~(\ref{Re_s}) holds
if
\begin{equation}
 \lambda\tau < \pi/2 \;.
 \label{uniformCondition}
\end{equation}
Long-time dynamics of the solution of Eq.~(\ref{h_evol}) is
governed by the zero(s) of Eq.~(\ref{char_eq_uniform}) with the {\em
largest} real part. In particular, for $\lambda\tau \le 1/e$, the
zero with the largest real part is purely real, hence no sustained
oscillations occur [Fig.~\ref{uniformModeEvol}(a)]. For
$1/e<\lambda\tau<\pi/2$, all zeros have imaginary parts (including
the ones with the largest real part) and are arranged symmetrically
about the real axis. This results in persistent oscillations that do
not diverge so long as condition (\ref{uniformCondition}) is
satisfied, as shown in Fig.~\ref{uniformModeEvol}(b).
The first pair of zeros to acquire positive real parts are the two
with smallest imaginary parts. Once the product $\lambda\tau$ fails
to satisfy the condition in Eq.~(\ref{uniformCondition}), the
oscillation amplitude grows in time [Fig.~\ref{uniformModeEvol}(c)].
Specific time series for $\langle h^2(t)\rangle$ are shown in
Fig.~\ref{uniformModeFluctEvol}, where the real parts of solutions
have become positive for delays $\tau = 1.60$ and $2.00$ but remain
negative for the rest.

\begin{figure}[t]
\centering
\includegraphics[scale=0.4]{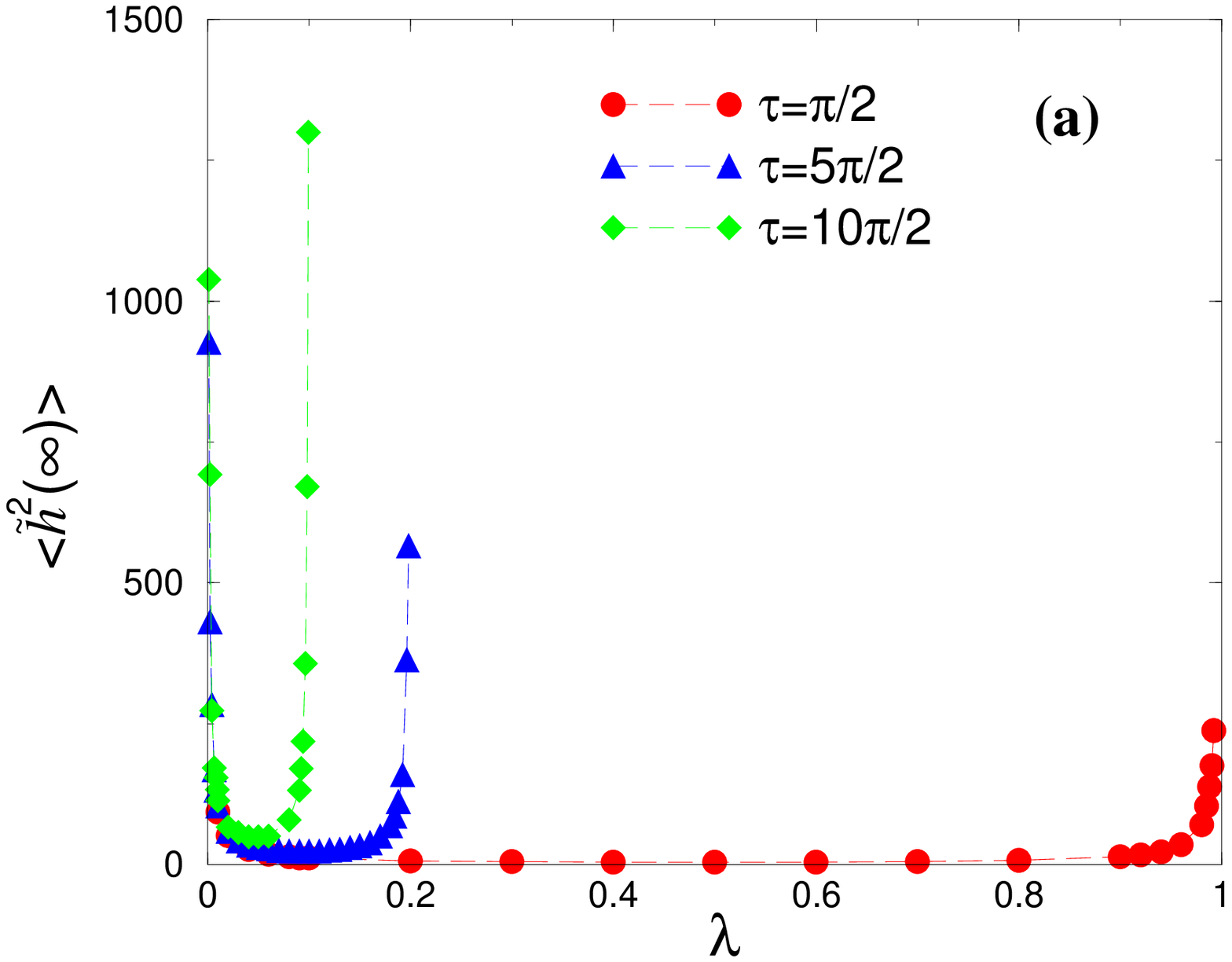}
\hspace*{1cm}
\includegraphics[scale=0.4]{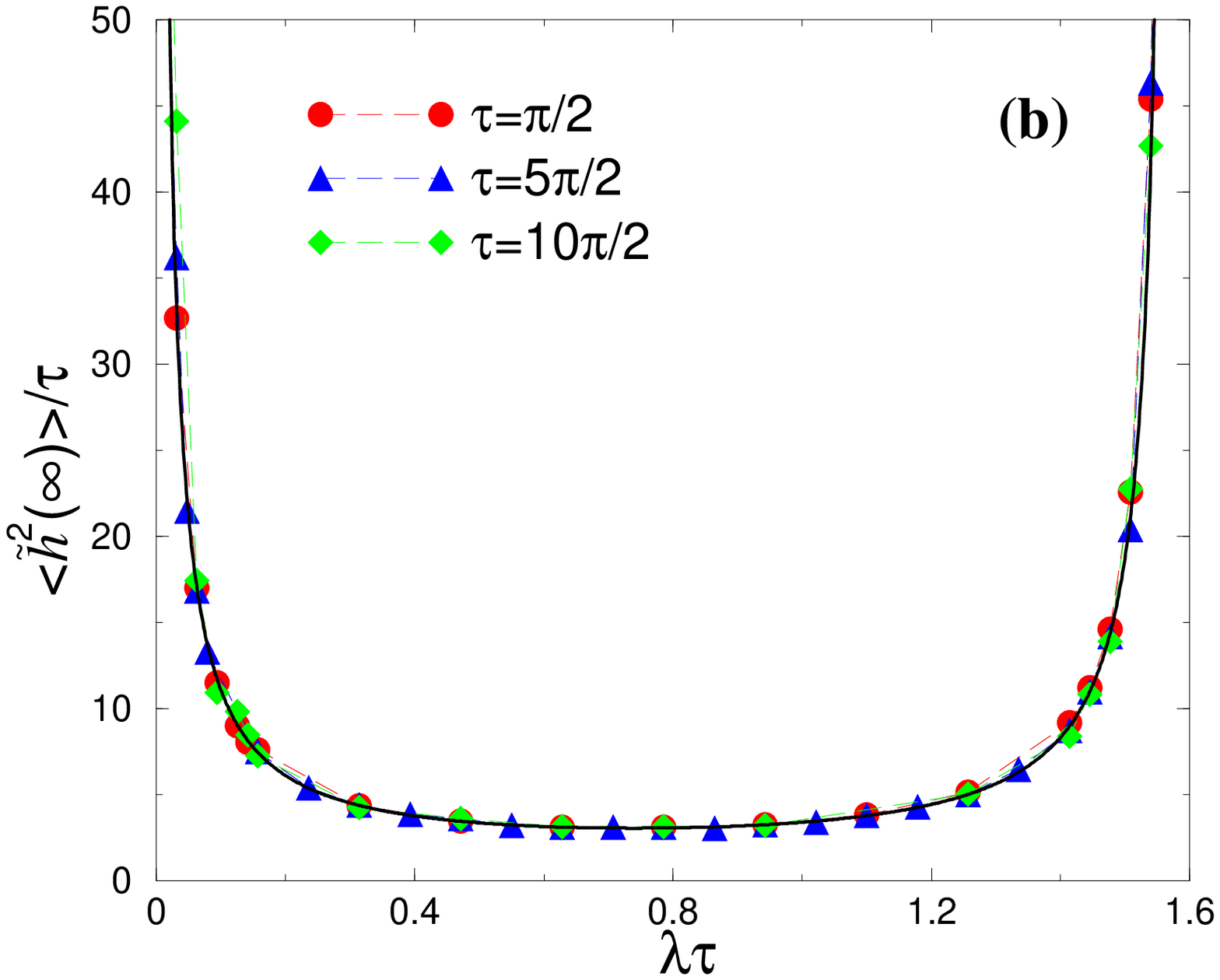}
\caption{(Color online)
(a) Steady-state fluctuations of an individual mode as a function of $\lambda$
obtained by numerical integration of Eq.~(\ref{h_evol}) for several delays with $D$$=$$1$ and $\Delta t$$=$$0.01$.
(b) Scaled fluctuations of an individual mode and the analytic scaling function Eq.~(\ref{uniformExact}) (solid curve).}
\label{uniformScaling}
\end{figure}

To obtain the general scaling form of the fluctuations in the
stationary state, we define $z_{\alpha}\equiv\tau s_{\alpha}$
($\alpha=1,2,\ldots$). One can easily see that the new variables
$z_{\alpha}$ are the corresponding solutions of the scaled
characteristic equation,
\begin{equation}
z + \lambda\tau e^{-z} = 0 \;,
\end{equation}
and hence can {\em only} depend on $\lambda\tau$, i.e. $z_{\alpha}=z_{\alpha}(\lambda\tau)$.
Thus,
\begin{equation}
s_{\alpha}(\lambda,\tau)= \frac{1}{\tau} z_{\alpha}(\lambda\tau)\;.
\end{equation}
Substituting this into Eq.~(\ref{h2_st}) yields
\begin{equation}
\langle\tilde h^2(\infty)\rangle = D\tau f(\lambda\tau)\;,
\label{uniformScaling_eq}
\end{equation}
where
\begin{equation}
f(\lambda\tau)= \sum_{\alpha, \beta}\frac{-2}{(1 + z_\alpha)(1 + z_\beta)(z_\alpha + z_\beta)}
\end{equation}
is the scaling function.
This scaling [Eq.~(\ref{uniformScaling_eq})] is illustrated by plotting $\langle\tilde h^2(\infty)\rangle/\tau$ vs $\lambda\tau$, fully collapsing the data for different $\tau$ values (with fixed noise
intensity $D$) [Fig.~\ref{uniformScaling}].

As mentioned earlier, Eq.~(\ref{h_evol}) has an exact solution for
the stationary-state variance obtained by K\"uchler and Mensch
\cite{Kuechler_SSR1992} (briefly reviewed in
Appendix~\ref{appendix_ScalingFunction}), providing an exact form
for the scaling function
\begin{equation}
f(\lambda\tau) = \frac{1 + \sin(\lambda\tau)}{\lambda\tau\cos(\lambda\tau)} \;.
\label{uniformExact}
\end{equation}
The asymptotic behavior of the scaling function near the singular points, $\lambda\tau=0$ and $\lambda\tau=\pi/2$, can be immediately extracted from the exact solution given by Eq.~(\ref{uniformExact}) (see also Ref.~\cite{Hod} for a more generalizable method),
\begin{equation}
f(\lambda\tau) \simeq \begin{cases}
\displaystyle\frac{1}{\lambda\tau} & ~~0 < \lambda\tau \ll 1 \\
\displaystyle\frac{4}{\pi(\pi/2 - \lambda\tau)} & ~~\displaystyle 0 < \frac{\pi}{2} - \lambda\tau \ll 1 \;.
\end{cases}
\end{equation}
The scaling function $f(x)$ ($x \equiv \lambda\tau$) is clearly {\em
non-monotonic}; it exhibits a single minimum, at approximately $x^*
\approx 0.739$ with $f^* = f(x^*) \approx 3.06$, found through
numerical minimization of Eq.~(\ref{uniformExact}). The immediate
message of the above result is rather interesting: For a single
stochastic variable governed by Eq.~(\ref{h_evol}) with a nonzero
delay, there is an optimal value of the ``relaxation" coefficient,
$\lambda^* = x^*/\tau$, at which point the stationary-state
fluctuations attain their minimum value
$\langle\tilde{h}^2(\infty)\rangle = D\tau f^* \approx 3.06 D\tau$.
This is in stark contrast with the zero-delay case (the standard
Ornstein-Uhlenbeck process \cite{Gardiner_1985}) where $\langle
\tilde{h}^2(\infty)\rangle = D/\lambda$, i.e., the stationary-state
fluctuation is a monotonically decreasing function of the relaxation
coefficient.

\subsection{Implications for Coordination in Unweighted Networks}

Since the eigenvectors of the Laplacian are orthogonal for symmetric couplings, the width can be expressed as the sum of the fluctuations for all non-uniform modes
\begin{equation}
\langle w^2(\infty)\rangle = \frac{1}{N}\sum_{k = 1}^{N - 1}\langle\tilde h_k^2(\infty)\rangle =
\frac{D\tau}{N}\sum_{k = 1}^{N - 1}f(\lambda_k\tau) \;,
\label{widthUniformDelay}
\end{equation}
where $\lambda_k$ is the eigenvalue of the $k$th mode.
Thus, condition (\ref{uniformCondition}) must be satisfied for every $k > 0$ mode for synchronizability, or equivalently \cite{consensus_threshold},
\begin{equation}
\lambda_{max}\tau < \frac{\pi}{2} \;.
\label{uniformNetworkCondition}
\end{equation}
The above exact delay threshold for synchronizability has some profound consequences for unweighted networks.
Here, the coupling matrix is identical to the adjacency matrix, $C_{ij} = A_{ij}$, and the bounds and the scaling properties of the extreme eigenvalues of the network Laplacian are well known.
In particular \cite{Fiedler_1973,Anderson_1985},
\begin{equation}
\frac{N}{N-1} k_{\rm max} \leq \lambda_{\rm max} \leq 2k_{\rm max} \;,
\end{equation}
where $k_{\rm max}$ is the maximum node degree in the network [i.e., $\langle\lambda_{\rm max}\rangle = {\cal O}(\langle k_{\rm max}\rangle)$].
Thus, $\tau k_{\rm max} < \pi/4$ is sufficient for synchronizibility \cite{consensus_threshold}, while $\tau k_{\rm max} > \pi/2$ leads to the breakdown of synchronization with certainty.
These inequalities imply that even a single (outlier) node with a sufficiently large degree can destroy synchronization or coordination in unweighted networks (regardless of the general trend, if any, of the tail of the degree distribution).
Naturally, network realizations selected from an ensemble of random graphs with a power-law tailed degree distribution typically have large hubs, making them rather vulnerable to intrinsic network delays \cite{Saber_IEEE2004,Saber_IEEE2007}.
For example, Barab\'asi-Albert (BA) \cite{Barab_sci,BarabREV} and uncorrelated \cite{Boguna_EPJB2004,Catanzaro_PRE2005} scale-free (SF) networks with structural degree cut-off (yielding $\lambda_{\rm max} \sim k_{\rm max} \sim N^{1/2}$) and similarly, SF network ensembles with natural cut-off (exhibting $\lambda_{\rm max} \sim k_{\rm max} \sim N^{1/(\gamma - 1)}$) for $N \gg 1$ \cite{MendesREV,Boguna_EPJB2004}), are particularly vulnerable.
Thus, for any fixed delay, increasing the size of scale free networks will eventually lead to the violation of condition (\ref{uniformNetworkCondition}), and in turn, to the breakdown of synchronization.
In contrast, the typical largest degree (hence the largest eigenvalue of the Laplacian) grows much slower in Erd\H{o}s-R\'enyi (ER) random graphs \cite{ErdosRenyiEvolution}, as $\lambda_{\rm max} \sim k_{\rm max} \sim \ln(N)$.

\begin{figure}[t]
\vspace{1.0truecm}
\centering
\includegraphics[scale=0.6]{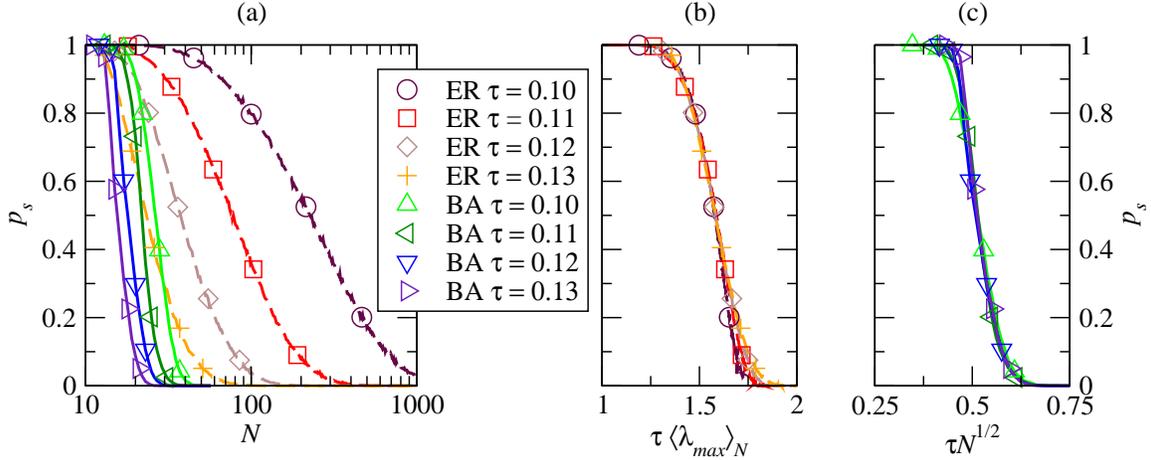}
\caption{(Color online) The fraction of synchronizable networks $p_s(\tau,N)$ taken from
ensembles of $10^4$ random constructions of ER and BA networks with $\langle k\rangle \approx 6$.
(a) $p_s$ vs. $N$. (b) and (c) are scaled plots of the same data according to Eq.~(\ref{fracSynchUniformScaling}),
for ER and BA networks, respectively.}
\label{uniformDelayFracSynch}
\end{figure}

To illustrate the above finite-size dependence, we define the fraction of synchronizable networks $p_{\rm s}(\tau,N)$, which is equivalent to the probability that a randomly chosen realization of a network ensemble satisfies $\lambda_{\rm max}$$<$$\pi/2\tau$.
Thus, $p_{\rm s}(\tau,N)=P_{N}^{<}(\pi/2\tau)$, where $P_{N}^{<}(x)$ is the cumulative probability distribution of the largest eigenvalue of the network Laplacian.
In Fig.~\ref{uniformDelayFracSynch}, we show the fraction of synchronizable networks for BA and ER network ensembles by employing direct numerical diagonalization of the corresponding network Laplacians and evaluating condition (\ref{uniformNetworkCondition}) for each realization.
For $N \gg 1$ the cumulative distribution for the largest eigenvalue exhibits the asymptotic scaling $P_{N}^{<}(x) \sim \phi(x/\langle \lambda_{\rm max}(N)\rangle)$ \cite{Kim_PRL2007}.
Thus, the fraction of synchronizable networks should scale as
\begin{equation}
p_{\rm s}(\tau,N)=P_{N}^{<}(\pi/2\tau)\sim\phi(\pi/2\tau\langle\lambda_{\rm max}(N)\rangle) = \psi(\tau\langle \lambda_{\rm max}(N)\rangle) \;.
\label{fracSynchUniformScaling}
\end{equation}
In Figs.~\ref{uniformDelayFracSynch} (b) and (c), we demonstrate the above scaling for ER and BA networks, respectively.

Since the scaling function is known exactly [Eq.~(\ref{uniformExact})], the eigenmode decomposition [Eq.~(\ref{widthUniformDelay})] allows one to evaluate the stationary width for an arbitrary network with a single uniform time delay by utilizing numerical diagonalization of the network Laplacian
\begin{equation}
\langle w^2(\infty)\rangle
= \frac{D\tau}{N}\sum_{k = 1}^{N - 1}f(\lambda_k\tau)
= \frac{D\tau}{N}\sum_{k = 1}^{N - 1}\frac{1 + \sin(\lambda_k\tau)}{\lambda_k\tau\cos(\lambda_k\tau)} \;.
\label{uniformWidthExact}
\end{equation}
The optimal (minimal) width occurs when all eigenvalues of the
Laplacian are degenerate so that the couplings and/or delay can be
tuned to the minimum of Eq.~(\ref{uniformExact}). For each mode in
Eq.~(\ref{uniformWidthExact}), such degeneracy is present in the
case of a fully-connected network with uniform couplings, optimized
to $C_{ij} = x^*/N\tau$ ($i \neq j$) and $C_{ii} = 0$. For general
networks, better synchronization can be achieved when the eigenvalue
spectrum is narrow relative to the range of synchronizability so
that most eigenvalues can fall near the minimum of
Eq.~(\ref{uniformExact}). We have not investigated in detail how to
achieve a narrow spectrum, but strategies for doing so by tuning
coupling strengths or adding/removing links have been explored by
others \cite{Nishikawa_PRE2006, Nishikawa_2009}.

\subsection{Scaling, Optimization, and Trade-offs in Networks with Uniform Delays}

\begin{figure}[t]
\vspace{1.0truecm}
\centering
\includegraphics[scale=0.6]{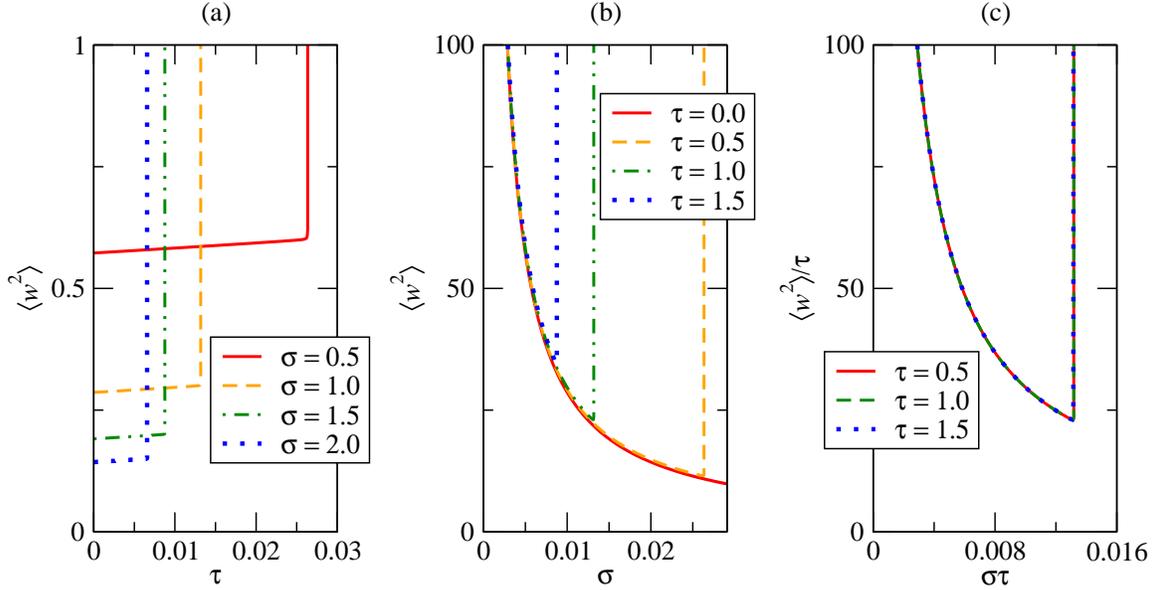}
\caption{(Color online) Stationary-state widths obtained through numerical diagonalization and utilizing Eq.~(\ref{uniformWidthExact}) for a typical BA network with $N = 100$
(a) for several coupling strengths,
(b) for several delays, and
(c) scaled so that the nonzero delay curves
collapse.}
\label{uniformWidthsVersusSigma}
\end{figure}

With the knowledge of the scaling function in Eq.~(\ref{uniformWidthExact}), one also immediately obtains the width for the case of an arbitrary but uniform effective coupling strength $\sigma$, where $C_{ij} = \sigma A_{ij}$.
The effective coupling strength can now be tuned for optimal synchronization.
However, there is a trade-off between how well the network synchronizes and the range over which it is synchronizable.
When the eigenvalue spectum is not narrow, diminishing the couplings uniformly in order to satisfy Eq.~(\ref{uniformNetworkCondition}) may cause small eigenvalues to be pushed farther up the left divergence of the scaling function.
Figure \ref{uniformWidthsVersusSigma}(a) shows this trade-off in uniform reweighting ($C_{ij} \rightarrow \sigma C_{ij}$).
The monotonicity of these widths means that the uniform delay should
{\em always} be minimized to obtain the best synchronization. The
same conclusion can be drawn from
Fig.~\ref{uniformWidthsVersusSigma}(b), which shows that networks
synchronize better and do not become unsynchronizable until greater
link strengths when the delay $\tau$ is minimized. Because globally
reweighting the coupling strengths corresponds to a uniform scaling
of the eigenvalues, we can define the width of a network by a
scaling function $F(\sigma\tau)$ (see
Fig.~\ref{uniformWidthsVersusSigma}(c))
\begin{equation}
\langle w^2(\infty)\rangle_{\sigma,\tau}
= \frac{D\tau}{N}\sum_{k = 1}^{N - 1} f(\sigma\lambda_k\tau)
= D\tau F(\sigma\tau).
\label{w2_scaling}
\end{equation}
Fluctuations from small eigenvalues dominate other contributions to the width for small $\sigma\tau$, hence the optimal value occurs near the end of the synchronizable region, where the network fails to meet condition (\ref{uniformNetworkCondition}).

\begin{figure}[t]
\vspace{1.0truecm}
\centering
\includegraphics[scale=0.6]{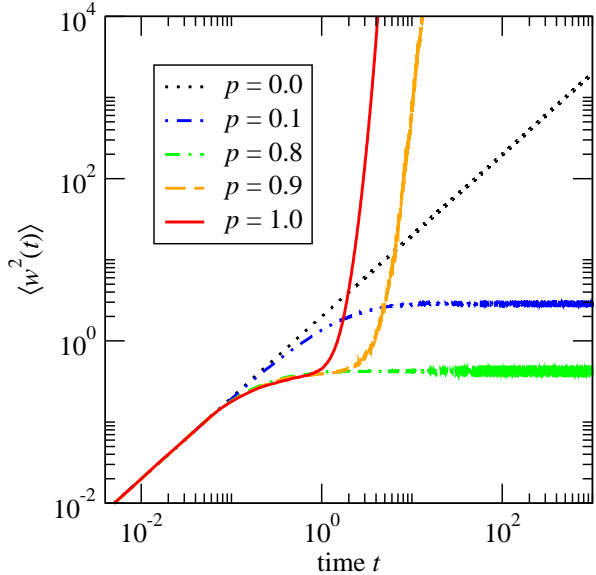}
\caption{(Color online) Time evolution of the width obtained by numerically
integrating Eq.~(\ref{h_evol_comm}) with $D$$=$$1$, $\Delta t$$=
$$0.005$, and averaged over $10^3$ realizations of noise for several
communication rates $p$ on a BA network of size $N$$=$$100$ and
average degree $\langle k\rangle$$=$$6$ with
$\tau\lambda_{max}$$=$$1.2\times\pi/2$.} \label{probEvol}
\end{figure}

As an alternative to varying the (effective) uniform coupling
strength $\sigma$, consider a scenario where the frequency (or rate)
of communication is controlled for each node according to
\begin{equation}
\partial_t h_i(t) = -p_i(t)\sum_{j=1}^N A_{ij}[h_i(t - \tau) - h_j(t - \tau)] + \eta_i(t) \;.
\label{h_evol_comm}
\end{equation}
In the above scheme, $p_i(t)$ is a binary stochastic variable for
each node, such that at each discretized time step, $p_i(t) = 1$
with probability $p$ and $p_i(t) = 0$ with probability $1 - p$ (for
simplicity, we employ uniform communication rates). The local
network neighborhood remains fixed, while nodes communicate with
their neighbors only at rate $p$ at each time step.
As an application for trade-off, consider a system governed by the
above equations and stressed by large delays, where local pairwise
communications at rate $p$$=$$1$ would yield unsynchronizability,
i.e., $\tau\lambda_{max}$$>$$\pi/2$ (see Fig.~\ref{probEvol}).
The width diverges for one of two reasons: either communication is
too frequent and the system fails to satisfy condition
(\ref{uniformNetworkCondition}), or there is no synchronization
($p$$=$$0$) and the system is overcome by noise. However, the
divergence of the width is faster in the former, accelerated by
overcorrections made by each node due to the delay. With an
appropriate reduction in the communication rate, the width reaches a
finite steady state, recovering synchronizability, as can be seen in
Fig.~\ref{probEvol}. Decreasing the frequency of communication can
counter-intuitively allow a network to become synchronizable for
delays and couplings that would otherwise cause the width to
diverge.

\subsection{Coordination and Scaling in Weighted Networks}

For the case of uniform delays, we compare two cases: networks with
weights that have been normalized locally by node degree and
networks with weights that are globally uniform. The couplings for
local weighting are defined as $C_{ij} = \sigma A_{ij}/k_i$ (a
common weighting scheme in generalized synchronization problems
\cite{Arenas_PhysRep2008}), while for uniform couplings $C_{ij} =
\sigma A_{ij}/\langle k\rangle$. In turn, the weighted (or
normalized) Laplacian becomes $\Gamma = \sigma K^{-1}L$ where $K$ is
the diagonal matrix with node degrees on its diagonal, $K_{ij} =
\delta_{ij}k_i$, and $L$ is the graph Laplacian, $L_{ij}
\equiv\delta_{ij}\sum_{l}A_{il} - A_{ij} =\delta_{ij}k_i - A_{ij}$.
Similarly, for uniform couplings, the corresponding Laplacian
becomes $\Gamma = \sigma\langle k\rangle^{-1}L$. Note that the
overall coupling strength (communication cost) is the same in both
cases, $\sigma\sum_{ij}A_{ij}/k_i = \sigma\sum_{ij}A_{ij}/\langle k
\rangle = \sigma N$.

\begin{figure}[t]
\vspace{1.5truecm}
\centering
\includegraphics[scale=0.6]{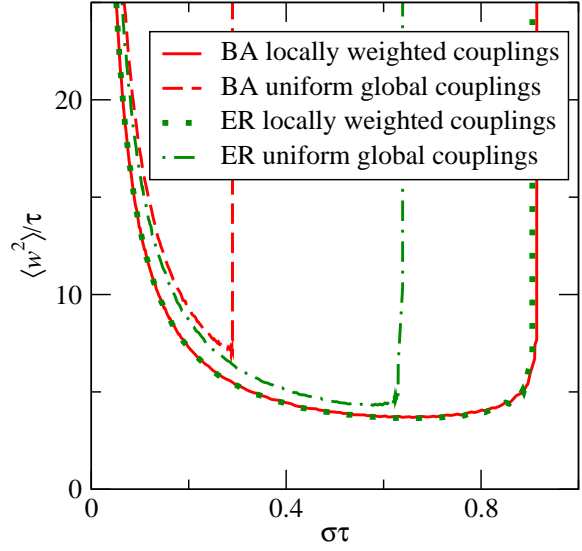}
\caption{(Color online) Scaled widths simulated with $\Delta t$$=$$0.01$ of a typical BA and a typical ER network,
each of size $N$$=$$100$ and with $\langle k\rangle$$=$$6$.}
\label{uniformDelayNormWidths}
\end{figure}

In the locally-weighted case, the eigenvalue spectrum of $K^{-1}L$
is known to be confined within the interval $[0, 2]$
\cite{networkSpectra}, so any network of this class will be
synchronizable, provided $\sigma\tau < \pi/4$. With globally uniform
weighting,  the increase of $\lambda_{max}$ with $N$ will lead to
fewer synchronizable networks as $N$ grows (holding $\langle
k\rangle$ constant).
\begin{figure}[t]
\vspace{1.0truecm}
\centering
\includegraphics[scale=0.6]{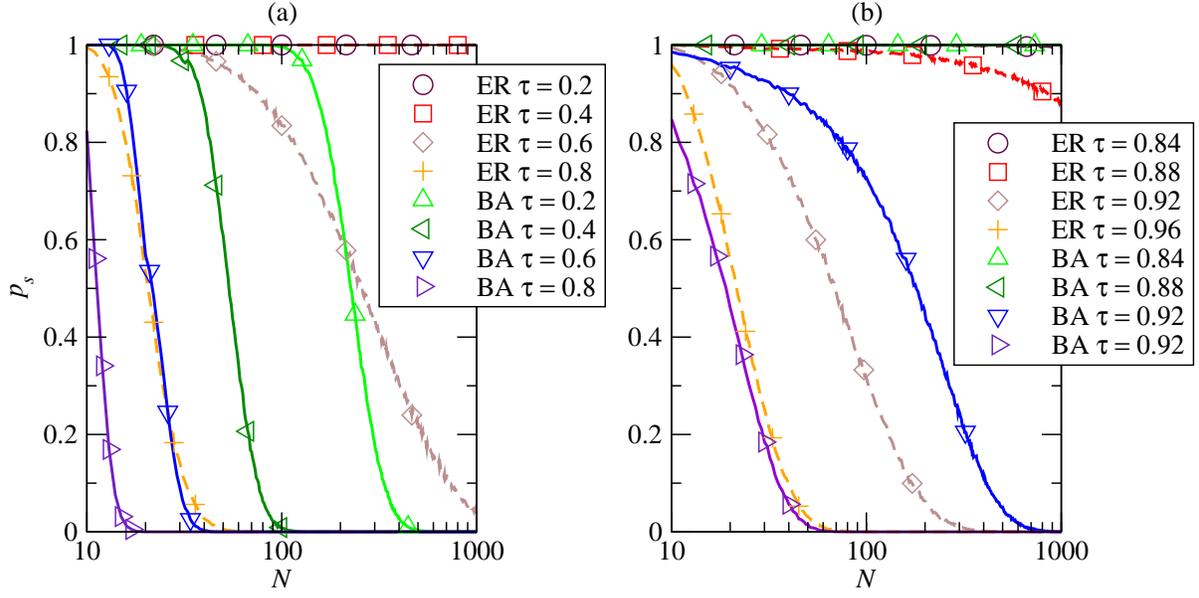}
\caption{(Color online) Fraction of synchronizable networks for (a) uniform global
weights and (b) local weights for the same ensemble of networks used
in Fig.~\ref{uniformDelayFracSynch}. }
\label{uniformDelayNormFracSynch}
\end{figure}
Figure \ref{uniformDelayNormFracSynch}(a) shows that it is more
likely for an ER network to be synchronizable than a BA network of
the same size $N$ when the couplings are weighted uniformly by
$\langle k\rangle$ (with all ER networks remaining synchronizable
over the range of $N$ for the two smallest delays). However, this is
not always the case when couplings are weighted locally by node
degree (Fig.~\ref{uniformDelayNormFracSynch}(b)), although nearly
all of these networks remain synchronizable over the delays in
Fig.~\ref{uniformDelayNormFracSynch}(a). The behavior of the width
for typical networks is shown in Fig.~\ref{uniformDelayNormWidths}
to compare the effects of these two normalizations.
In both the BA and ER case, synchronization is better and is
maintained for longer delays when the coupling strengths are
weighted locally by node degree.

\section{Multiple Time Delays}

To generalize the basic model, we now allow for a distinction in transmission and processing time delays.
In this case, Eq.~(\ref{diffEqGeneral}) becomes
\begin{equation}
\partial_t h_i(t) = -\sum_j C_{ij}[h_i(t - \tau_{\rm o}) - h_j(t - \tau_{\rm o} - \tau_{\rm tr})] + \eta_i(t)
\label{diffEqTrans}
\end{equation}
where the local delay $\tau_{\rm o}$ and the transmission delay
$\tau_{\rm tr}$ are the same for all nodes and links, respectively.
Although the synchronizability condition and steady state width
cannot be determined in a closed form for arbitrary networks as is
the case of Eq.~(\ref{diffEqUniform}), focusing on special cases
does offer insight.

\subsection{Fully-Connected Networks}

\begin{figure}[t]
\vspace{1.0truecm}
\centering
\includegraphics[scale=0.6]{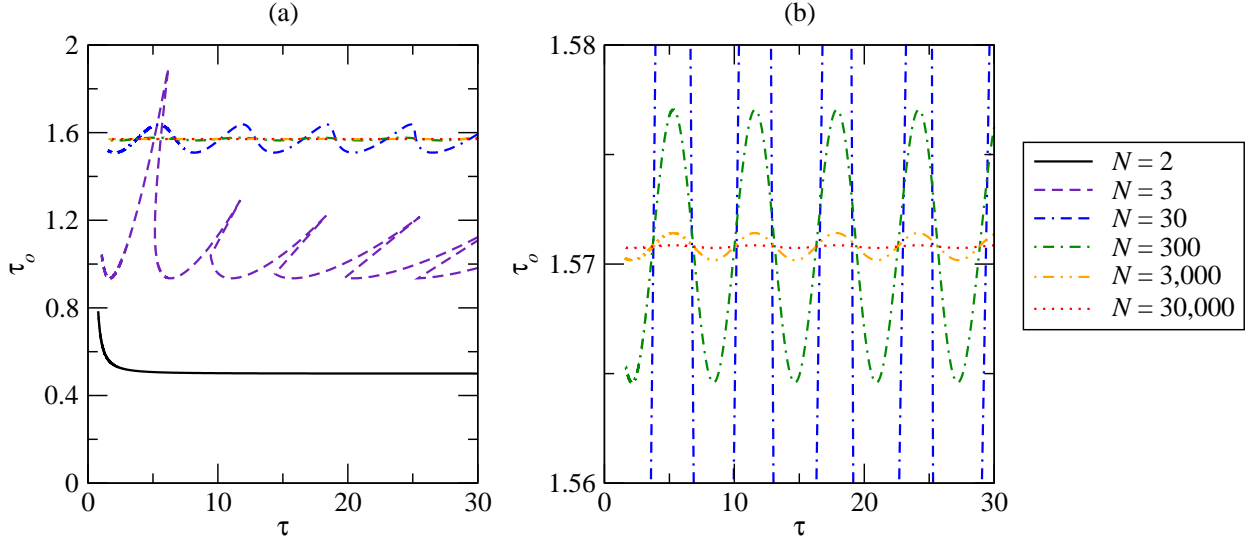}
\caption{(Color online) Phase diagram (synchronization boundary) for fully-connected networks with uniform
coupling strength $\sigma/(N-1)$ in the $(\tau_{\rm o},\tau)$ plane.
(Here, without loss of generality due to scaling, we used $\sigma$$=$$1$.)
(a) All system sizes (for $N$$\geq$$300$ they are essentially indistinguishable from the $N$$=$30,000 case
at these scales);
(b) system sizes $N$$=$$30$, 300, 3,000, 30,000 in an enlarged region for visibility.
With the exception of the analytically solvable case of $N$$=$$2$ \cite{HuntPLA}, the synchronization boundaries,
corresponding to stability limits, were obtained from the analysis of the zeros of Eq.~(\ref{fullyConnectedCharEq}).}
\label{fullyConnectedTwoDelayBoundary}
\vspace*{-0.20truecm}
\end{figure}

Consider the case of a fully-connected network of size $N$ with
uniform link strengths $\sigma$, where the local state variables
evolve according to
\begin{align}
\partial_t h_i(t) &  = -\frac{\sigma}{N - 1}\sum_{j \ne i}[h_i(t - \tau_o) - h_j(t - \tau)] + \eta_i(t)
                     = -\frac{\sigma}{N - 1}\sum_{j \ne i}[h_i(t - \gamma\tau) - h_j(t - \tau)] + \eta_i(t)
\nonumber \\
 & = -\frac{\sigma}{N - 1}\sum_{j \ne i}[h_i(t - \tau) - h_j(t - \tau)] + \sigma h_i(t - \tau) - \sigma h_i(t - \gamma\tau) + \eta_i(t)
\nonumber \\
 & = -\frac{\sigma}{N - 1}\sum_j \Gamma_{ij}h_j(t - \tau) + \sigma h_i(t - \tau) - \sigma h_i(t - \gamma\tau) + \eta_i(t)
\label{diffEqFC}
\end{align}
where $\tau \equiv \tau_{\rm o} + \tau_{\rm tr}$, $\gamma \equiv
\tau_{\rm o}/\tau$ and $\Gamma_{ij} = \delta_{ij}N - 1$. Normalizing
the global coupling with $1/(N-1)$ assures that the coupling cost
per node remains constant and the region of synchronization remains
finite in the limit of $N \to \infty$. Using the fact that the graph
Laplacian of the complete graphs has a single, nonzero eigenvalue
$N$ [which is ($N$$-$$1$)-fold degenerate], each non-uniform mode
(associated with fluctuations about the mean) obeys
\begin{equation}
\partial_t \tilde h(t) = - \sigma \tilde h(t - \gamma\tau) - \frac{\sigma}{N-1}\tilde h(t - \tau) + \tilde{\eta}(t) \;.
\label{diffEqFCMode}
\end{equation}
As in the case of uniform delays, we perform a Laplace transform on the deterministic part to obtain the characteristic polynomial and equation,
\begin{equation}
g(s) \equiv  s + \frac{\sigma}{N-1}e^{-\tau s} + \sigma e^{-\gamma\tau s} = 0 \;.
\label{fullyConnectedCharEq}
\end{equation}
Note that for $N = 2$, the region of stability/synchronizability can
be obtained analytically \cite{HuntPLA}, and for completeness we
show it in Fig.~\ref{fullyConnectedTwoDelayBoundary} in the
$(\tau_{\rm o},\tau)$ plane. In this simple case of two coupled
nodes, the synchronization boundary is monotonic, and the local
delay is dominant: There is no singularity (for any finite
$\tau_{\rm tr}$) as long as $\sigma\tau_o < 1/2$ \cite{HuntPLA},
while for any $\tau_{\rm tr}$, there is a sufficiently large
$\tau_o$ resulting in the breakdown of synchronization.

For $N \ge 3$, the phase diagram (region of synchronizability) can
be obtained numerically by tracking the zeros of the characteristic
equation Eq.~(\ref{fullyConnectedCharEq}) (i.e., identifying when
their real parts switch sign) shown in
Fig.~\ref{fullyConnectedTwoDelayBoundary}. Note that keeping track
of infinitely many complex zeros of the characteristic equations
would be an insurmountable task. Instead, in order to identify the
stability boundary of the system, one only needs to know whether
{\em all} solutions have negative real parts. This test can be done
by employing Cauchy's argument principle
\cite{Hoefener2011,Luz_JCAM1996} (see Appendix~\ref{appendix_Cauchy}
for details).
Similar to the $N$$=$$2$ case, the local delay is always dominant,
i.e., there are critical values of $\sigma\tau_{\rm o}$ above/below
which the system is unsynchronizable/synchronizable for any
$\tau_{\rm tr}$. [These critical values approach $\pi/2$ as $N \to
\infty$, since in this case Eqs.~(\ref{diffEqFCMode}) and
(\ref{fullyConnectedCharEq}) reduce to the familiar forms of
Eqs.~(\ref{h_evol}) and (\ref{char_eq_uniform}), respectively, with
the known analytic threshold.] The behavior with the overall delay
$\tau=\tau_{\rm o} + \tau_{\rm tr}$, however, is more subtle: There
is a range of $\tau_o$ where varying $\tau$ yields reentrant
behavior with alternating synchronizable and unsynchronizable
regions (as can be seen by considering suitably chosen horizontal
cuts for fixed $\tau_{\rm o}$ in
Fig.~\ref{fullyConnectedTwoDelayBoundary}).
Thus, in this region (for fixed local delays $\tau_{\rm o}$),
stabilization of the system can also be achieved by {\em increasing}
the transmission delays.

\begin{figure}[t]
\vspace{1.6truecm}
\centering
\includegraphics[scale=0.6]{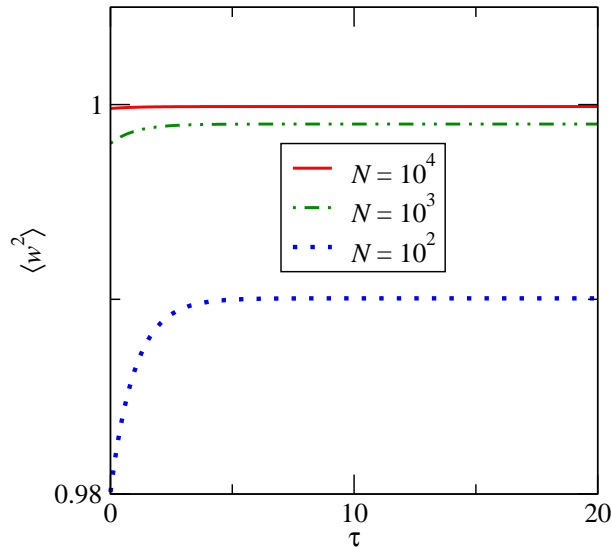}
\caption{(Color online) Analytic results for stationary-state widths for
fully-connected networks of several sizes for the special case
$\gamma$$=$$0$ [Eq.~(\ref{w2_FCexact})].  Here, $D$$=$$1$ and
$\sigma$$=$$1$.}
\label{fullyConnectedNoBaseWidths}
\end{figure}

In the special case $\gamma = 0$, the network is always
synchronizable for all $N$ and the width can be obtained exactly
(see Appendix \ref{App_special}.b),
\begin{equation}
\langle w^2(\infty)\rangle =
\frac{1}{N}\sum_{k = 1}^{N - 1}\langle\tilde h_k^2(\infty)\rangle =
\frac{D(N - 1)}{N}\frac{\alpha + \frac{\sigma}{N-1}\sinh(\alpha\tau)}{\alpha[\sigma + \frac{\sigma}{N-1}\cosh(\alpha\tau)]}
\label{w2_FCexact}
\end{equation}
with $\alpha=\sigma\sqrt{1-1/(N-1)^2}$, as shown in
Fig.~\ref{fullyConnectedNoBaseWidths}.
For $\tau = \tau_{\rm tr} \rightarrow \infty$, the above expression becomes
\begin{equation}
\langle w^2(\infty)\rangle = \frac{D(N - 1)}{N}\frac{1}{\sigma\sqrt{1 - 1/(N - 1)^2}} \;.
\end{equation}

\subsection{Locally Weighted Networks}

\begin{figure}[t]
\vspace{1.5truecm}
\centering
\includegraphics[scale=0.6]{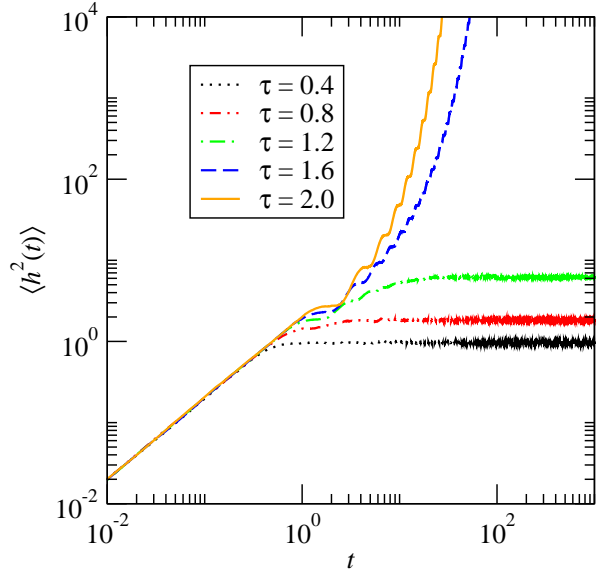}
\caption{(Color online) Time series of the fluctuations of a single mode for several delays obtained from numerical
integration of Eq.~(\ref{diffEqNormMode}) with $\gamma$$=$$0.5$,
$\lambda$$=$$1.8$, $D$$=$$1$, and $\Delta t$$=$$0.01$, averaged over $10^3$ realizations of the noise ensemble.}
\label{normalModeEvol}
\vspace*{-0.20truecm}
\end{figure}

\begin{figure}[t]
\vspace{1.0truecm}
\centering
\includegraphics[scale=0.6]{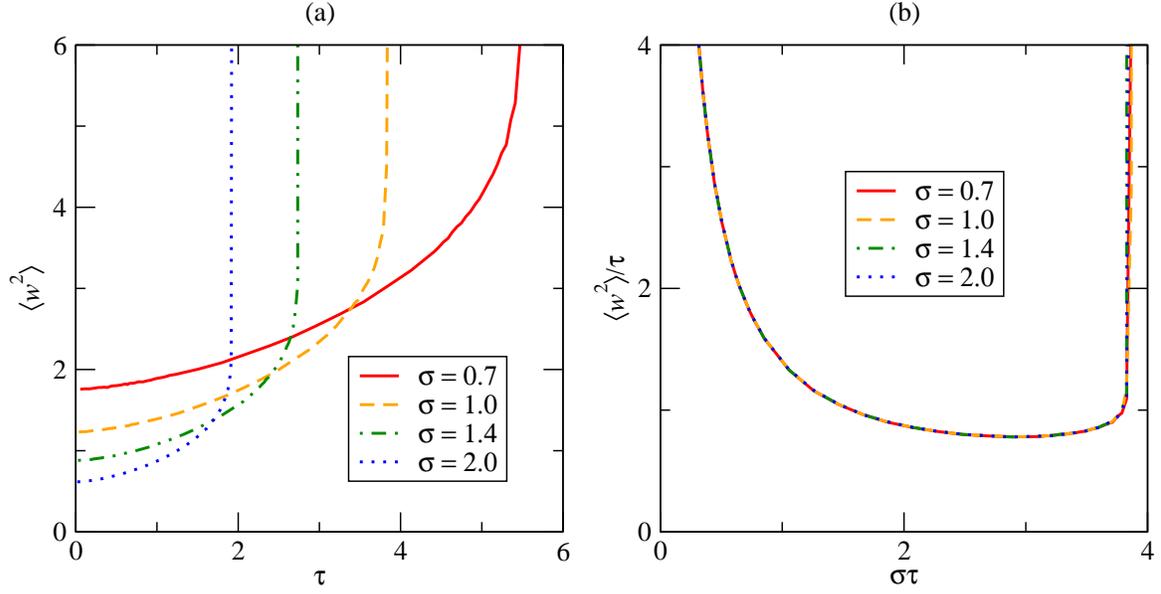}
\caption{(Color online)
Comparison of (a) the widths and (b) the scaled widths for
several coupling strengths $\sigma$ on a typical locally weighted BA
network of size $N$$=$$100$ and $\langle k\rangle \approx 6$ for
$\gamma$$=$$0.2$; simulated with $D$$=$$1$ and $\Delta t$$=$$0.001$.}
\label{normedTwoDelayWidths}
\end{figure}

Now we consider Eq.~(\ref{diffEqTrans}) with specific locally
weighted couplings (already utilized for uniform local time delays
in Sec.II.D), $C_{ij}=\sigma A_{ij}/k_i$. The set of differential
equations then have the form
\begin{align}
\partial_t h_i(t) & = -\frac{\sigma}{k_i}\sum_j A_{ij} [h_i(t - \gamma\tau) - h_j(t - \tau)] + \eta_i(t) \nonumber \\
 & = -\frac{\sigma}{k_i}\sum_j L_{ij}h_j(t - \tau) + \sigma h_i(t - \tau) - \sigma h_i(t - \gamma\tau) + \eta_i(t) \nonumber \\
 & = -\sigma\sum_j\Gamma_{ij}h_j(t - \tau) + \sigma h_i(t - \tau) - \sigma h_i(t - \gamma\tau) + \eta_i(t) \;,
\label{diffEqNorm}
\end{align}
where $\sigma$ controls the coupling strength and $\Gamma = K^{-1}L$
is now the locally weighted network Laplacian ($K_{ij} =
\delta_{ij}k_i$, and $L_{ij} = \delta_{ij}\sum_{l}A_{il} - A_{ij} =
\delta_{ij}k_i - A_{ij}$). Diagonalization yields
\begin{equation}
\partial_t\tilde h_k(t) = \sigma(1 - \lambda_k)\tilde h_k(t - \tau) - \sigma\tilde h_k(t - \gamma\tau) + \tilde\eta_k(t)
\label{diffEqNormMode}
\end{equation}
where $\lambda_k$ is the eigenvalue of the $k$th mode of the
normalized graph Laplacian $K^{-1}L$.
Figure \ref{normalModeEvol}
shows the evolutions of a particular mode with delays on either side
of the critical delay.
The characterisitic equation for the $k$th mode is then
\begin{equation}
g_{k}(s) = s + \sigma(\lambda_k - 1)e^{-\tau s} + \sigma e^{-\gamma\tau s} = 0 \;.
\label{char_eq_gk}
\end{equation}
Defining the new scaled variable $z = \tau s$, this equation becomes
\begin{equation}
z + (\sigma\tau)(\lambda_k - 1)e^{-z} + (\sigma\tau)e^{-\gamma z} = 0 \;.
\label{normCharEq}
\end{equation}
Hence, the solutions of the original characteristic equation depends
on $\sigma$ and $\tau$ in the form of
$s_{k\alpha}=\tau^{-1}z_{k\alpha}(\sigma\tau)$. Although the scaling
function of the width in the case of locally normalized couplings
with two time delays cannot be expressed in a closed form, the
general scaling behavior is identical to Eq.~(\ref{w2_scaling}) [as
follows from the formal solution shown in
Appendix~\ref{appendix_UniformMode}, Eq.~(\ref{hh_ss})], i.e.,
$\langle w^2(\infty)\rangle_{\sigma,\tau} = D\tau F(\sigma\tau)$.
The corresponding scaling behavior and scaling collapse, obtained
from numerical integration of Eq.~(\ref{diffEqNorm}), are shown in
Fig.~\ref{normedTwoDelayWidths}.

\begin{figure}[t]
\vspace*{1.0truecm}
\centering
\includegraphics[scale=0.6]{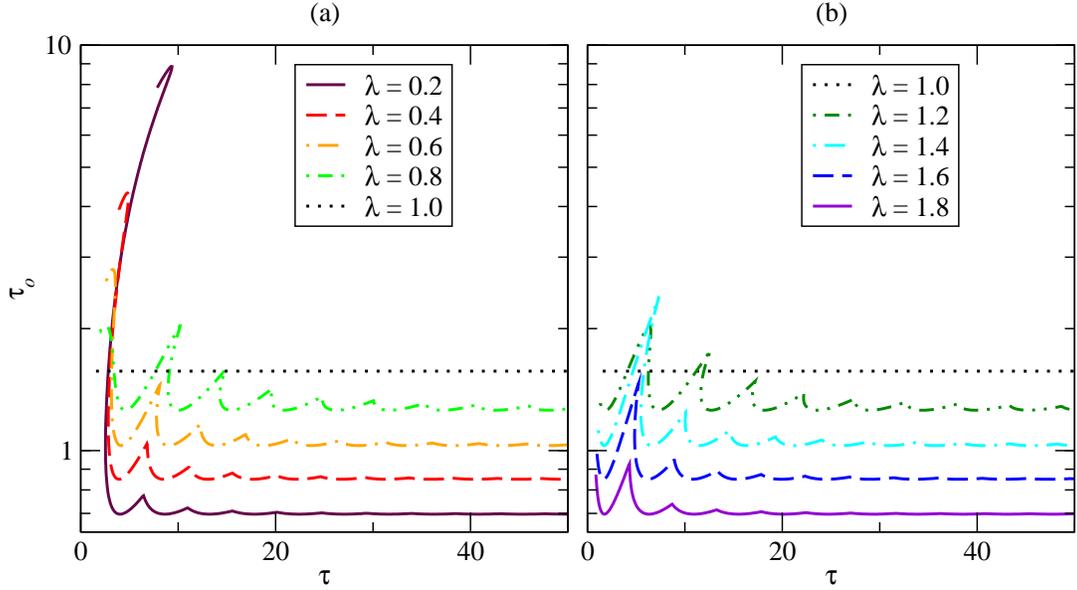}
\caption{(Color online) Synchronization boundaries for several modes with (a)
$\lambda_k \le 1$ and (b) $\lambda_k \ge 1$ of a weighted network,
obeying Eq.~(\ref{diffEqNormMode}) and determined by analyzing the zeros of Eq. (\ref{normCharEq}). }
\label{normalizedBoundaryModes}
\end{figure}

\begin{figure}[t]
\vspace{1.0truecm}
\centering
\includegraphics[scale=0.6]{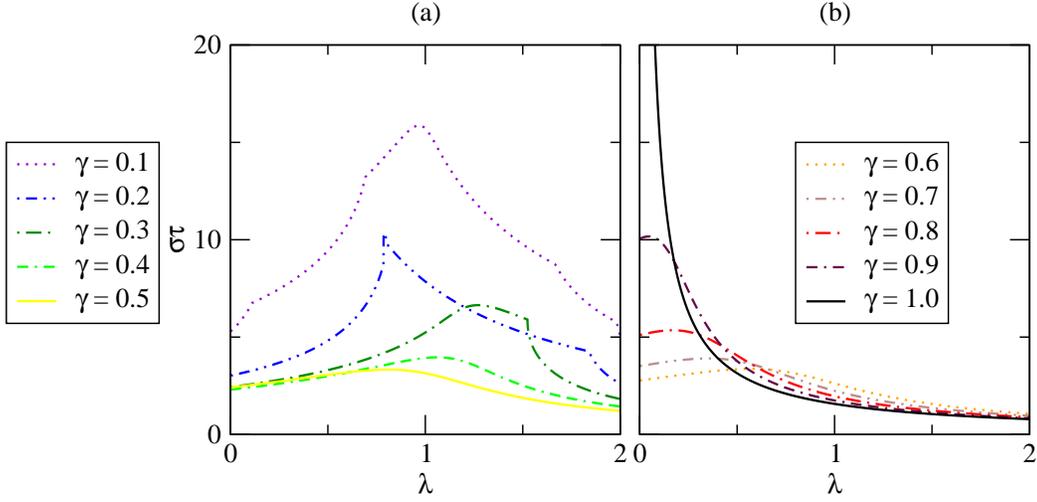}
\caption{(Color online)
Synchronization boundaries determined by analyzing the zeros of Eq.~(\ref{normCharEq}) for various delay ratios
$\gamma$, shown separately for (a) $\gamma \le 0.5$ and (b) $\gamma  \ge 0.6$.}
\label{normalizedBoundaryVsLambda}
\end{figure}

\begin{figure}[t]
\vspace{1.0truecm}
\centering
\includegraphics[scale=0.6]{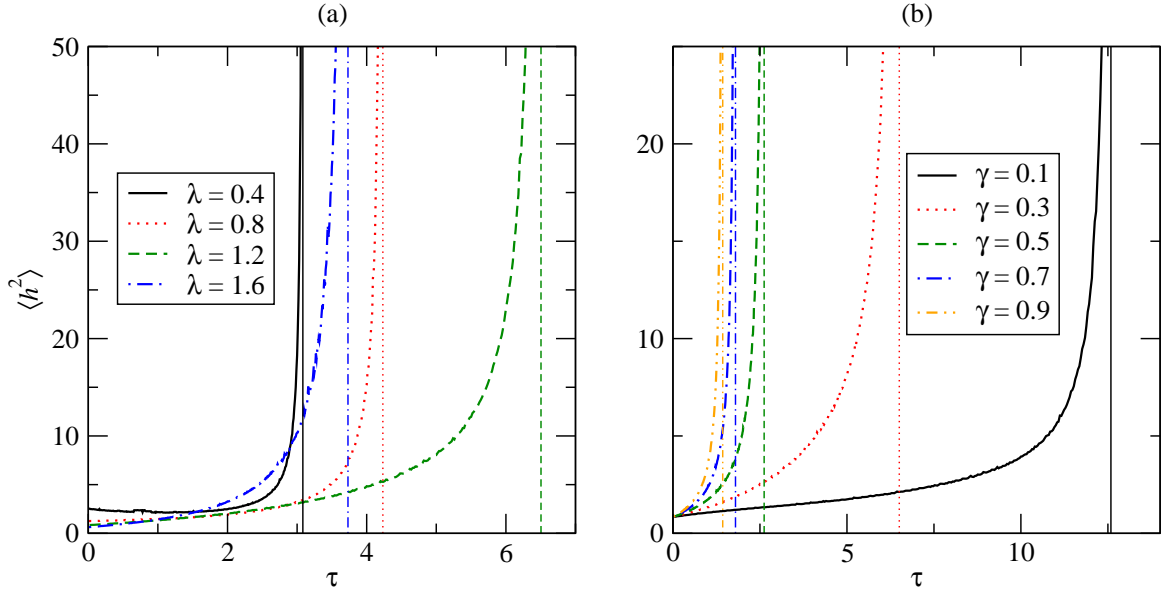}
\caption{(Color online) Width contributions for (a) several modes with $\gamma =
0.3$ and (b) several delay ratios with $\lambda = 1.2$, found by
numerically integrating Eq.~(\ref{diffEqNormMode}) with $D$$=$$1$
and $\sigma$$=$$1$. The vertical lines
correspond to the stability limits obtained from the analyses of the
zeros of Eq.~(\ref{char_eq_gk}) with the same $\lambda$.}
\label{normModeWidths}
\vspace*{-0.20truecm}
\end{figure}

The stability/synchronization boundary was again determined by
employing Cauchy's argument principle \cite{Hoefener2011,Luz_JCAM1996},
applied separately for each mode (Appendix~\ref{appendix_Cauchy}).
Figure \ref{normalizedBoundaryModes} shows the most important
eigenvalues to determine synchronizability: the greatest restriction
to the critical delay $\tau_c = (\tau_{\rm o} + \tau_{\rm tr})_c$
for a given $\gamma$ belongs to either the smallest or largest
eigenvalues.
An alternative presentation is given in
Fig.~\ref{normalizedBoundaryVsLambda}, which shows that it is not
always the same eigenvalue that consistently limits
synchronizability for all values of $\gamma$; rather it is the
eigenvalue that falls on the lowest point on the boundary curve.
The contributions of a few example modes to the width are shown in
Fig.~\ref{normModeWidths}(a). Note that the order of divergences is
not the same as the ordered eigenvalues, in accordance with
Fig.~\ref{normalizedBoundaryModes}. The contributions of a single
mode for various values of $\gamma$ is shown in
Fig.~\ref{normModeWidths}(b). Since it is $\tau_{\rm o}$ that has a
greater impact on whether or not a network can synchronize, larger
total delays $\tau$ are tolerated for smaller $\gamma$ since more of
the delay comes from transmission.
Because of the great sensitivity of $\langle
h^2\rangle$ on $\Delta t$ near the divergence for longer delays, an
adaptive algorithm was implemented, which would halve $\Delta t$
until consecutive runs agreed within 1\%.

\begin{figure}[t]
\vspace{1.0truecm}
\centering
\includegraphics[scale=0.6]{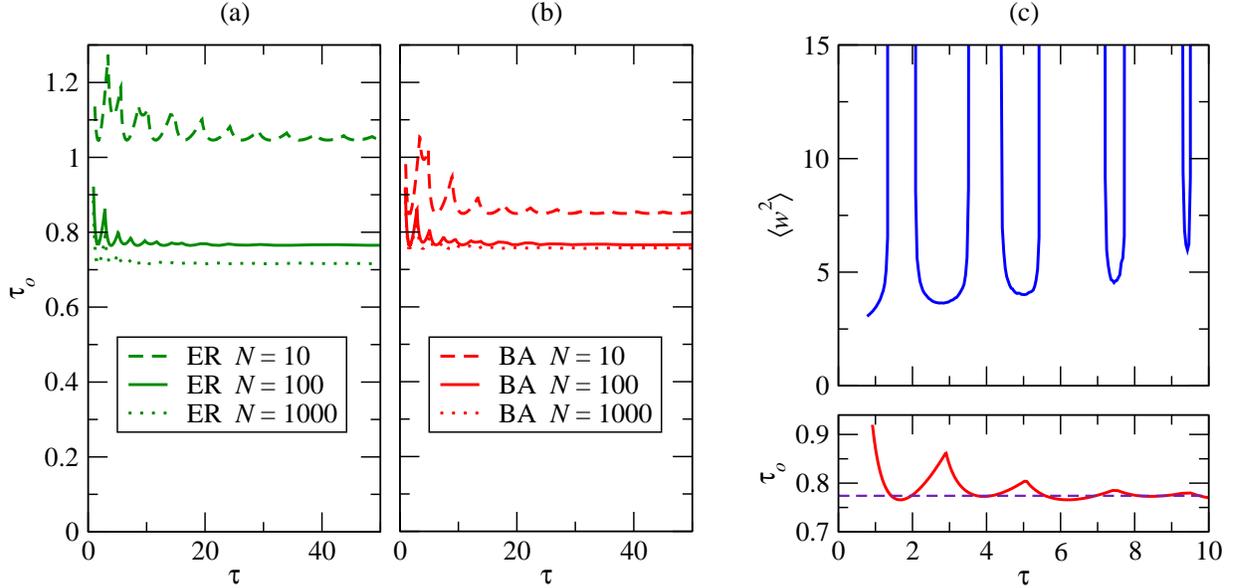}
\caption{(Color online) Synchronization boundaries for typical (a) ER and (b) BA
networks of several sizes with locally weighted couplings. The
boundaries are found by numerical diagonalization and examining each
mode through Eq.~(\ref{char_eq_gk}). (c) Widths along a slice of constant $\tau_o$$=$$0.77$ for the same
$N$$=$$100$ BA network used in (b).  For stability comparison, the boundary is shown below with the slice indicated.}
\label{normedTwoDelayBoundary}
\end{figure}

\begin{figure}[t]
\vspace{1.75truecm}
\centering
\includegraphics[scale=0.6]{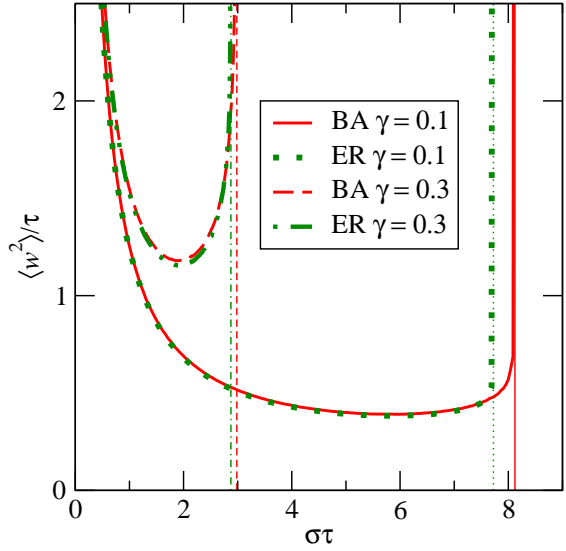}
\caption{(Color online) The scaling functions of a typical locally weighted BA
network and a typical ER network for two delay ratios, with both
networks of size $N$$=$$100$, found by numerically integrating
Eq.~(\ref{diffEqNorm}) with $D$$=$$1$ and $\Delta t$$=$$ 0.001$. The
vertical lines correspond to the stability limits obtained from the
analyses of the zeros of Eqs.~(\ref{normCharEq}).}
\label{normWidths100}
\end{figure}

With this understanding of the underlying modes, let us return to
synchronization of the entire system. Incorporating all relevant
eigenvalues results in the synchronization boundary shown in
Fig.~\ref{normedTwoDelayBoundary}(a) for several representative
networks.
The cut for a carefully chosen local delay in
Fig.~\ref{normedTwoDelayBoundary}(b) shows the previously mentioned
reentrant behavior as the transmission delay is increased. Note that
the optimal width within each synchronizable region worsens with
larger delay, so that while synchronizability can be recovered with
increasing $\tau_{tr}$, better synchronization is possible by
decreasing $\tau_{tr}$. To compare the contribution of modes within
the synchronizable regime, consider again the two topologies of BA
and ER graphs. For fixed $\gamma$, Fig.~\ref{normWidths100} shows
that a BA graph remains synchronizable for larger delays than a ER
graph when the link strengths are weighted by node degree.
However, the ER graph synchronizes slightly better for the majority of the time that it is synchronizable.
Here it is not the topology but the ratio $\gamma$ that has the most drastic effect.

When $\gamma < 1$, the mode corresponding to $\lambda_0 = 0$ includes self-interaction terms and has the critical delay
\begin{equation}
\tau_c(\lambda = 0) = \frac{\pi}{1 + \gamma}\frac{1}{\displaystyle\left|\cos\left(\pi\frac{1 - \gamma}{1 + \gamma}\right)\right|}.
\end{equation}
While the uniform mode does not contribute to the width because
$\bar h$ is removed from the state of the network (see Appendix
\ref{appendix_UniformMode}), a diverging mean can introduce
egregious truncation errors into the numerical integration if $\bar
h$ diverges exponentially while the width remains finite.
Fortunately, this can be avoided by simulating the network in the
subspace lacking the zero mode by removing the mean from each time
slice. Since the uniform mode is not allowed propagate, it does not
cause any problem with finite precision. The locations of the zeros'
real parts for Eq.~(\ref{normCharEq}) are tracked again using
Cauchy's argument principle (see Appendix~\ref{appendix_Cauchy}).

\subsection{Arbitrary Couplings and Multiple Delays}

\begin{figure}[t]
\vspace{1.5truecm}
\centering
\includegraphics[scale=0.6]{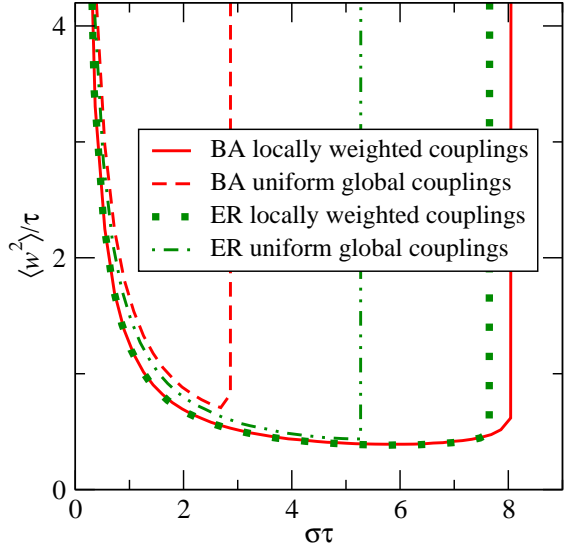}
\caption{(Color online) Scaled width curves for a typical BA network compared to
those of a typical ER network of size $N = 100$ with $\langle
k\rangle \approx 6$ and $D = 1$, determined by numerically
integrating Eq.~(\ref{diffEqTrans}) for the two types of coupling
schemes with $\gamma = 0.1$ and $\Delta t = 0.01$.}
\label{normalizations}
\end{figure}

When there are multiple time delays involved in the synchronization
or coordination process, in general, one cannot diagonalize the
underlying system of coupled equations. This happens to be the case
for the scenario with two types of time delay
[Eq.~(\ref{diffEqTrans})] on {\em unweighted} (or globally weighted)
graphs (as opposed to specific locally-weighted ones discussed in
Sec.~III.B). First, we briefly present a generally applicable method
to determine the region of synchronizability/stability
computationally \cite{Hoefener2011,Luz_JCAM1996}. For arbitrary
couplings $C_{ij}$, the deterministic part of
Eq.~(\ref{diffEqTrans}) (from which one can extract the
characteristic equation) becomes
\begin{equation}
\partial_t h_i(t) = -C_{i}h_i(t - \tau_{\rm o})  + \sum_j C_{ij} h_j(t - \tau) \;,
\end{equation}
where $C_i$$=$$\sum_{l}C_{il}$ and $\tau=\tau_{\rm o} + \tau_{\rm tr}$. After Laplace transform,
these equations become
\begin{equation}
s\hat{h}_i(s) = -C_{i}\hat{h}_i(s) e^{-s\tau_{\rm o}}  + \sum_j C_{ij} \hat{h}_j(s) e^{-s\tau} \;,
\end{equation}
or equivalently,
\begin{equation}
\sum_{j}\left( s\delta_{ij} + C_{i}\delta_{ij} e^{-s\tau_{\rm o}} - C_{ij} e^{-s\tau} \right)\hat{h}_j(s)  = 0 \;.
\end{equation}
Hence, non-trivial solutions of the above system of equations require
\begin{equation}
\det M(s)  = 0 \;,
\label{generalCharEq}
\end{equation}
where
\begin{equation}
M_{ij}(s) = s\delta_{ij} + C_{i} e^{-s\tau_{\rm o}} \delta_{ij} - C_{ij} e^{-s\tau} \;.
\label{generalCharEqdef}
\end{equation}
Stability or synchronizability requires that ${\rm Re}(s)$$<$$0$ for
{\em all} solutions of the above (transcendental) characteristic
equation [Eq.~(\ref{generalCharEq})]. To identify the stability
boundary of this coupled system, one does not need to know and
determine the (infinitely many) complex solutions of the
characteristic equation, but only whether all solutions have
negative real parts. To test that, one again can employ the argument
principle \cite{Hoefener2011,Luz_JCAM1996}
(Appendix~\ref{appendix_Cauchy}). Note that the above method can be
immediately generalized to arbitrary heterogeneous (local and
transmission) time delays. To compare synchronizability with locally
weighted couplings of the same cost [Eq.~(\ref{diffEqNorm})], here,
we considered $C_{ij}=\sigma A_{ij}/\langle k\rangle$. The results
are shown in Fig.~\ref{normalizations}. The synchronization boundary
was determined using the above scheme, while the width was obtain by
numerically integrating Eq.~(\ref{diffEqTrans}). Not only does local
reweighting of the coupling strength improve synchronization, but it
also {\em extends} the region of synchronizability.

\section{Summary}

Through our investigations we have explored the impact and interplay
of time delays, network structure, and coupling strength on
synchronization and coordination in complex interconnected systems.
Here, we considered only linear couplings, already yielding a rather
rich phase diagram and response.
While nonlinear effects are crucial in all real-life applications
\cite{Huberman_IEEE1991,Chen_EPL2008,Chen_PRE2009,Chen_PLOS2011},
linearization and stability analysis about the synchronized state
yields equations analogous to the ones considered here
\cite{Jadbabaie_IEEE2010,Strogatz_PRE2003}. Hence, the detailed
analysis of the linear problems can provide some insights to the
complex phase diagrams and response of nonlinear problems.

For a single uniform local delay, the synchronizability of a network
is governed solely by the largest eigenvalue and the time delay.
This result links the presence of larger hubs to the vulnerability
of the system becoming unstable at smaller delays. The quality of
synchronization within the stable regime is described by the width,
which can be enumerated exactly for arbitrary symmetric couplings,
provided the spectrum is known. We have also established the
boundaries of the region of synhronizability in terms of the delay
and the overall coupling strength (associated with communication
rate) and provided the general scaling behavior of the width inside
this region. Our results underscored the importance of the interplay
of stochastic effects, network connections, and time delays, in that
how ``less" (in terms of local communication efforts) can be ``more"
efficient (in terms of global performance).

For more general schemes with multiple time delays, we have shown
how stability analysis in general delay differential equations can
be applied to ascertain the synchronizability of a network.
For cases where, at least in principle, eigenmode decomposition is
possible, we have identified the general scaling behavior of the
width within the synchronizable regime. However, in these cases
it is not always the same eigenvalue that determines stability for all $\gamma$.
In the
non-monotonic nature of the scaling function, we see that there is a
fundamental limit to how well a network can synchronize in the
presence of noise. In the case when transmission and reaction are
two independent and significant sources of delay, there is an
additional parameter for tuning: the ratio of local delay to the
total delay. By fixing the local delay and cutting across different
values of the ratio, there is the possibility that the network will
enter into and emerge from synchronizable regions. Understanding
these influences can guide network design in order to maintain and
optimize synchronization by balancing the trade-offs in internodal
communication and local processing.

\section*{Acknowledgments}
We thank A. Asztalos for comments on the manuscript. This
work was supported in part by DTRA Award No. HDTRA1-09-1-0049, by
the Army Research Laboratory under Cooperative Agreement Number
W911NF-09-2-0053, by the Office of Naval Research Grant No.
N00014-09-1-0607, and by NSF Grant No. DMR-1246958. The views and conclusions contained in this
document are those of the authors and should not be interpreted as
representing the official policies, either expressed or implied, of
the Army Research Laboratory or the US Government.


\appendix

\section{Steady-state fluctuations for a single-variable stochastic delay equation}
\label{appendix_Fluctuations}

For a single (linearized) stochastic variable $h(t)$ with multiple
time delays $\{\tau_\omega\}_{\omega=1}^{\Omega}$ and
delta-correlated noise, one starts with the following general form
\begin{equation}
\partial_{t} h(t) = A_{0}h(t) + \sum_{\omega=1}^{\Omega} A_\omega h(t-\tau_\omega) + \eta(t) \;,
\label{noise_delay_eq}
\end{equation}
where $\langle\eta(t)\eta(t')\rangle$$=$$2D\delta(t-t')$. Formally,
the noise $\eta(t)$ plays the role of the inhomogeneous part of the
above inhomogeneous linear first-order differential equation.
Performing Laplace transformation
[$\hat{h}(s)=\int_{0}^{\infty}e^{-st}h(t)dt$], the characteristic
polynomial $g(s)$ (and the corresponding characteristic equation)
associated with the homogeneous (deterministic) part of the above
equation becomes
\begin{equation}
g(s) \equiv s - A_{0}  - \sum_{\omega=1}^{S} A_\omega e^{-\tau_{\omega}s} = 0\;.
\label{char_eq}
\end{equation}
For the initial condition $h(t) \equiv 0$ for $t \leq 0$ (which we
employ throughout this paper), we can easily obtain the Laplace
transformed Green's function of Eq.~(\ref{noise_delay_eq}) (i.e.,
the solution when $\eta(t)$ is replaced by $\delta(t-t')$), which
has the form
\begin{equation}
\hat{G}(s) = \frac{e^{-st'}}{g(s)} \;.
\label{G_s}
\end{equation}
Performing the inverse transform, one finds
\begin{equation}
G(t,t') = \frac{1}{2\pi i}  \int_{x_0-i\infty}^{x_0+i\infty} ds e^{st} \hat{G}(s)
= \frac{1}{2\pi i} \int_{x_0-i\infty}^{x_0+i\infty} ds \frac{e^{s(t-t')}}{g(s)}
= \Theta (t-t')\sum_\alpha\frac{e^{s_{\alpha}(t - t')}}{g^{'}(s_{\alpha})} \;,
\label{G_t}
\end{equation}
where $s_{\alpha}$ ($\alpha=1,2,\ldots$) are the zeros of the characteristic equation $g(s) = 0$ on the complex plane [Eq.~(\ref{char_eq})].
In the above inverse transform, the infinite line of integration is parallel to the imaginary axis ($s=x_0$) and is chosen to be to the right of all zeros of the characteristic polynomial in order to apply the residue theorem by closing the contour
with an infinite semicircle to the left of this line.
Note that the Green's function $G(t,t')$ depends only on the variable $t - t'$, reflecting the time translation symmetry of the problem.
Utilizing the Grenn's function, we can now formally write the general solution of
Eq.~(\ref{noise_delay_eq}) (for the same initial conditions) as
\begin{equation}
h(t) = \int_0^{\infty} dt'G(t,t')\eta(t') = \int_0^{t} dt'G(t,t')\eta(t')
= \int_0^t dt' \sum_\alpha\frac{e^{s_{\alpha}(t - t')}}{g^{'}(s_{\alpha})}\; \eta(t') \;.
\label{noise_delay_sol}
\end{equation}
For more general initial conditions, see Ref.~\cite{Bambi_JEDC2008}.

After averaging over the noise, one finds that the fluctuations of $h(t)$ are
\begin{eqnarray}
\langle h^2(t)\rangle  & = &
\left\langle\int_0^t dt' \eta(t')\sum_\alpha\frac{e^{s_{\alpha}(t - t')}}{g^{'}(s_{\alpha})}
\int_0^t dt'' \eta(t'')\sum_\beta\frac{e^{s_{\beta}(t - t'')}}{g^{'}(s_{\beta})}\right\rangle \nonumber \\
& = &
\int_0^t dt' \int_0^t dt'' \sum_\alpha\frac{e^{s_{\alpha}(t - t')}}{g^{'}(s_{\alpha})}
\sum_\beta\frac{e^{s_{\beta}(t - t'')}}{g^{'}(s_{\beta})} \langle\eta(t')\eta(t'')\rangle \nonumber \\
& = &
\sum_{\alpha,\beta} \frac{1}{g^{'}(s_{\alpha}) g^{'}(s_{\beta})}
\int_0^t dt' \int_0^t dt'' \; e^{s_{\alpha}(t - t')} e^{s_{\beta}(t - t'')} 2D\delta(t'-t'') \nonumber \\
& = &
\sum_{\alpha,\beta} \frac{2D}{g^{'}(s_{\alpha}) g^{'}(s_{\beta})}
\int_0^t dt' \; e^{(s_{\alpha} + s_{\beta})(t - t')} \nonumber \\
& = &
\sum_{\alpha,\beta} \frac{-2D( 1- e^{(s_{\alpha} + s_{\beta})t} )}{g^{'}(s_{\alpha}) g^{'}(s_{\beta})(s_{\alpha} + s_{\beta})} \;.
\label{h2_delay_sol}
\end{eqnarray}
As is explicit from the above equation, the zero with the largest
real part of all $s_{\alpha}$ governs the long-time behavior of the
stochastic variable $h(t)$. In particular, $h(t)$ reaches a
stationary limit distribution for $t\to\infty$ with a finite
variance if and only if ${\rm Re}(s_\alpha) < 0$ for all $\alpha$.
In this case,
\begin{equation}
\langle h^2(\infty)\rangle  = \sum_{\alpha,\beta}
\frac{-2D}{g^{'}(s_{\alpha}) g^{'}(s_{\beta})(s_{\alpha} + s_{\beta})} \;,
\end{equation}
otherwise it diverges exponentially with time.
Note that the condition for the existence of an asymptotic stationary limit distribution is the same as the one for the {\em stability} of the deterministic (homogeneous) part of Eq.~(\ref{noise_delay_eq}) about the $h_i=0$ fixed point \cite{Ruan_2006,Saber_IEEE2004}.

\section{Exact Scaling Functions for Time Delayed Stochastic Differential Equations}
\label{appendix_ScalingFunction}

K\"uchler and Mensch \cite{Kuechler_SSR1992} obtained the analytic
stationary-state autocorrelation function for the stochastic
delay-differential equation,
\begin{equation}
\partial_{t} h(t) = ah(t) + bh(t-\tau) + \eta(t) \;,
\label{KM_noise_delay_eq}
\end{equation}
with $\langle\eta(t)\eta(t')\rangle = 2D\delta(t-t')$. Special cases
(with suitably chosen coefficients $a$ and $b$) can be directly
utilized for two of our special cases in our network investigations.
Specifically, for ({\it i}) unweighted networks with symmetric
couplings and uniform local time delays with no transmission delays
and ({\it ii}) the case with only transmission delays on the
complete graph, to be discussed at the end of this Appendix, in
Appendix~\ref{App_special}.a and Appendix~\ref{App_special}.b,
respectively. Here, we briefly present an equivalent derivation of
their results, using the formalism used in our paper.

We define the {\em stationary-state} autocorrelation function as
\begin{equation}
K(t) = \langle  h(t')h(t'+t)\rangle \;,
\label{K_def}
\end{equation}
where it is implicitly assumed that $t' \to \infty$. From this
definition and the invariance under time translation in the
stationary state, it follows that the interpretation of the
autocorrelation function can be formally extended to $t$$<$$0$ by
\begin{equation}
K(t) = \langle  h(t')h(t'+t)\rangle = \langle h(t'+t)h(t')\rangle = \langle h(t')h(t'-t)\rangle = K(-t) \;,
\label{K_ext}
\end{equation}
and also,
\begin{equation}
\dot{K}(t)= -\dot{K}(-t) \;.
\label{K_dotext}
\end{equation}

As one would like to obtain a directly solvable equation of motion for the autocorrelation function, one must first find expressions for its time derivatives.
Employing the equation of motion for $h(t)$ [Eq.~(\ref{KM_noise_delay_eq})], we obtain for $t \geq 0$ that
\begin{eqnarray}
\dot{K}(t) & = & \partial_{t}K(t) = \partial_{t}\langle h(t')h(t'+t)\rangle = \langle h(t')\partial_{t}h(t'+t)\rangle =
\langle h(t')\{ ah(t'+t) + bh(t'+t-\tau) + \eta(t'+t)\} \rangle \nonumber \\
& = & a\langle h(t')h(t'+t)\rangle + b\langle h(t')h(t'+t-\tau) \rangle + \langle h(t')\eta(t'+t)\rangle \\
& = & aK(t) + bK(t-\tau) \;,
\label{K_dot}
\end{eqnarray}
where in the last step we used $\langle h(t')\eta(t'+t)\rangle=0$
(i.e., Ito's convention \cite{Gardiner_1985,vKampen_JSP1981}). The
above expression, combined with the (analytic) extension of the
autocorrelation function in Eq.~(\ref{K_ext}), yields the condition
\begin{equation}
\dot{K}(0) = aK(0) + bK(\tau) \;
\label{K_cond}
\end{equation}
in the limit of $t \to +0$. Differentiating Eq.~(\ref{K_dot}) again
with respect to $t$ and exploiting the properties of
Eqs.~(\ref{K_ext}) and (\ref{K_dotext}), we find
\begin{eqnarray}
\ddot{K}(t) & = & a\dot{K}(t) + b\dot{K}(t-\tau) = a\dot{K}(t) - b\dot{K}(\tau-t) \nonumber \\
& = & a\{ aK(t) + bK(t-\tau) \} - b\{ aK(\tau-t) + bK(-t) \} \nonumber \\
& = & a\{ aK(t) + bK(t-\tau) \} - b\{ aK(t-\tau) + bK(t) \} \nonumber \\
& = & (a^2 - b^2) K(t) \;.
\label{K_dot2}
\end{eqnarray}
Note that the reduction of the equation of motion of the
autocorrelation function to a second order ordinary differential
equation (with no delay) is a consequence of
Eq.~(\ref{KM_noise_delay_eq}) having only one delay time-scale. The
general solution of Eq.~(\ref{K_dot2}) can be written as
\begin{equation}
K(t) = A\cos(\omega t) + B\sin(\omega t)
\end{equation}
with $\omega = \sqrt{b^2 - a^2}$. From the definition of the
autocorrelation function in Eq.~(\ref{K_def}) and from some of the
basic properties of the Green's function (see Appendix
\ref{App_Green_prop} below for details), it also follows
\cite{Kuechler_SSR1992} that
\begin{equation}
\dot{K}(0) = \lim_{t\to 0}\partial_{t}\langle h(t')h(t'+t)\rangle = -D \;,
\label{K_cond1}
\end{equation}
and from Eq.~(\ref{K_cond}),
\begin{equation}
aK(0) + bK(\tau) = -D \;.
\label{K_cond2}
\end{equation}
Thus, the second order ordinary differential equation Eq.~(\ref{K_dot2}) with conditions Eqs.~(\ref{K_cond1}) and
(\ref{K_cond2}) can now be fully solved, yielding
\begin{equation}
A = K(0)= D\frac{-\omega + b\sin(\omega\tau)}{\omega[a + b\cos(\omega\tau)]} \;,
\label{K_A}
\end{equation}
and
\begin{equation}
B = \frac{\dot{K}(0)}{\omega} =  -\frac{D}{\omega}\;.
\label{K_B}
\end{equation}
Finally, the stationary-state variance of the stochastic variable governed by Eq.~(\ref{KM_noise_delay_eq}) can be written as
\begin{equation}
\langle h^2(t)\rangle =  \langle h(t)h(t)\rangle = K(0) =
D\frac{-\omega + b\sin(\omega\tau)}{\omega[a + b\cos(\omega\tau)]} \;.
\label{h2_exact}
\end{equation}
Following the aforementioned technical detours and details in Appendix~\ref{App_Green_prop}, we will
discuss in Appendix~\ref{App_special} the applications of the above result to obtain the scaling function of the fluctuations for the individual modes in specific networks.

\subsection{General properties of the autocorrelation function and the Green's Function}
\label{App_Green_prop}

From the definition of the autocorrelation function in Eq.~(\ref{K_def}) and of the Green's function in Eq.~(\ref{noise_delay_sol}), it follows that
\begin{eqnarray}
K(t) & = & \langle h(t')h(t'+t)\rangle  = \left\langle \int_0^{t'} du\; G(t',u)\eta(u)  \int_0^{t'+t} dv\; G(t'+t,v)\eta(v) \right\rangle \nonumber \\
     & = &   \int_0^{t'} du \int_0^{t'+t} dv \;G(t',u) G(t'+t,v)  \langle\eta(u)\eta(v)\rangle = 2D \int_0^{t'} du\; G(t',u)G(t'+t,u) \;,
\end{eqnarray}
and consequently
\begin{eqnarray}
\dot{K}(t) & = & \partial_{t}\langle h(t')h(t'+t)\rangle = 2D \partial_{t} \int_0^{t'} du\; G(t',u)G(t'+t,u) = 2D  \int_0^{t'} du\; G(t',u) \partial_{t} G(t'+t,u) \nonumber \\
           & = & 2D \int_0^{t'} du\; G(t',u) (-\partial_{u}) G(t'+t,u) = -2D \int_0^{t'} du\; G(t',u) \partial_{u} G(t'+t,u) \;.
\end{eqnarray}
Hence,
\begin{eqnarray}
\dot{K}(0) & = & -2D \lim_{t\to 0} \int_0^{t'} du\; G(t',u) \partial_{u} G(t'+t,u) = -2D \int_0^{t'} du\; G(t',u) \partial_{u} G(t',u) \\
           & = & -2D \int_0^{t'} du\; \partial_{u} \frac{G(t',u)^2}{2} = -D \{ G(t',t') - G(t',0) \} = -D\{ 1 - 0 \} = -D \; ,
\end{eqnarray}
where in the second term of the last expression above we now
explicitly exploited that $G(t',t') = 0$ and $G(t',0) \to 0$ as $t'
\to \infty$. The former can be seen by a segment-by-segment
integration and solution of Eq.~(\ref{KM_noise_delay_eq}) with a
delta source $\delta(t-t')$ in the intervals
$(t-t')\in[n\tau,(n+1)\tau]$, $n=0,1,2,\ldots$
\cite{Kuechler_SSR1992}; the solution in the $[0,\tau]$  interval is
particularly simple, $G(t,t')=\exp[a(t-t')]$. The latter property is
trivial in that the magnitude of the Green's function in the
stationary state has to decay for large arguments.

\subsection{Applications to Special Cases}
\label{App_special}

\subsubsection{Unweighted Symmetric Couplings with Uniform Local Delays}

For symmetric couplings $C_{ij}$ with uniform local delays, the
Laplacian $\Gamma_{ij} = \delta_{ij}\sum_{l} C_{il} - C_{ij}$ in
Eq.~(\ref{diffEqUniform}) can, in principle, be diagonalized. Each
mode is governed by Eq.~(\ref{hk_evol}), a special case of
Eq.~(\ref{KM_noise_delay_eq}) with $a = 0$, $b = -\lambda$, and
$\omega = |b| = \lambda$ ($\lambda$ being the eigenvalue of the
respective mode). From Eq.~(\ref{h2_exact}), the steady-state
variance of each mode then reduces to
\begin{equation}
\langle h^2(\infty)\rangle =
D\frac{1 + \sin(\lambda\tau)}{\lambda\cos(\lambda\tau)} =
D\tau\frac{1 + \sin(\lambda\tau)}{\lambda\tau\cos(\lambda\tau)} =
D\tau f(\lambda\tau) \;,
\label{h2_exact_uniform}
\end{equation}
yielding the analytic scaling function for each mode
\begin{equation}
f(x) = \frac{1 + \sin(x)}{x\cos(x)} \;,
\label{scaling_exact_uniform}
\end{equation}
with the scaling variable $x = \lambda\tau$.

\subsubsection{Complete Graphs with Only Uniform Transmission Delays}

The exact stationary-state variance of Eq.~(\ref{KM_noise_delay_eq})
can also be applied to complete graphs with global coupling
$\sigma$, which have no local delays but do have uniform
transmission delays, i.e., Eq.~(\ref{diffEqFCMode}) with $\gamma =
0$, translating to $a = -\sigma$, $b = -\sigma/(N-1)$ in
Eq.~(\ref{KM_noise_delay_eq}). The analytic expression from
Eq.~(\ref{h2_exact}) for the stationary-state variance for each
(non-uniform) mode becomes
\begin{equation}
\langle h^2(\infty)\rangle =
D\frac{\alpha + \frac{\sigma}{N-1}\sinh(\alpha\tau)}{\alpha[\sigma + \frac{\sigma}{N-1}\cosh(\alpha\tau)]} \;,
\label{h2_FCexact}
\end{equation}
with $\alpha=\sqrt{a^2 - b^2}=\sigma\sqrt{1 - 1/(N - 1)^2}$.

\section{Application of Cauchy's Argument Principle with Implementation}
\label{appendix_Cauchy}

For an arbitrary complex analytic function $F(z)$, the number of
zeros $N_C$ inside a closed contour $C$ (provided $F(z)$ has no
poles/singularities inside $C$) is given by Cauchy's argument
principle (see, e.g., Ref.~\cite{Krantz1999}):
\begin{equation}
N_C = \frac{1}{2\pi i}\oint_C\frac{F'(z)}{F(z)}dz = \frac{1}{2\pi}\Delta_C\arg F(z)\;,
\label{Cauchy}
\end{equation}
where $\Delta_C\arg F(z)$ is the winding number of $F(z)$ along the
closed contour $C$. The characteristic equations studied in this
paper can all be written as a sum of exponentials, hence there are
no singularities. To determine the stability boundary, we follow
Refs.~\cite{Hoefener2011,Luz_JCAM1996} and use Eq.~(\ref{Cauchy}) to
track the number of zeros of the characteristic equations with
positive real part (i.e., on the positive real half plane) by
substituting Eqs.~(\ref{char_eq_gk}) and (\ref{generalCharEq}) for
$F(z)$. We employed a numerical algorithm \cite{Hoefener2011} for
enumerating the winding number with adaptive step size. We restate
the method here: a step of size $h$ along the contour in the
direction $\hat\iota$ from $s$ to $s + h\hat\iota$ is accepted if
$\theta(s, s + (h/2)\hat\iota < \theta(s, s + h\hat\iota) < \epsilon
= 1$ where $\theta(s, s') \equiv |{\rm arg}(\det M(s)) - {\rm
arg}(\det M(s'))| {~\rm mod~} 2\pi$. The subsequent step size is
then $h \rightarrow \max\{2, \epsilon/\Delta\}$; unacceptable steps
are retried with $h \rightarrow h/2$. The winding number is the
count of the number of crossings of $\pi$ without a return in the
opposite direction.

We choose the contour so that it detects the first zero to cross the
imaginary axis and acquire a positive real part. Note that the mode
corresponding to the zero eigenvalue allows the solution $z$$=$$0$
for Eq.~(\ref{normCharEq}), so the zero at the origin
is ignored. This can be easily achieved by choosing the left edge of
the contour to be nonzero but still very small. This method can be
applied to any network structure with any delay scheme, provided the
approximate general behavior of the zeros is understood.

As an example, consider the simplest system of two coupled nodes with uniform delay, which has a critical delay of $\pi/4$.
\begin{figure}[t]
\vspace{1.0truecm}
\centering
\includegraphics[scale=0.4]{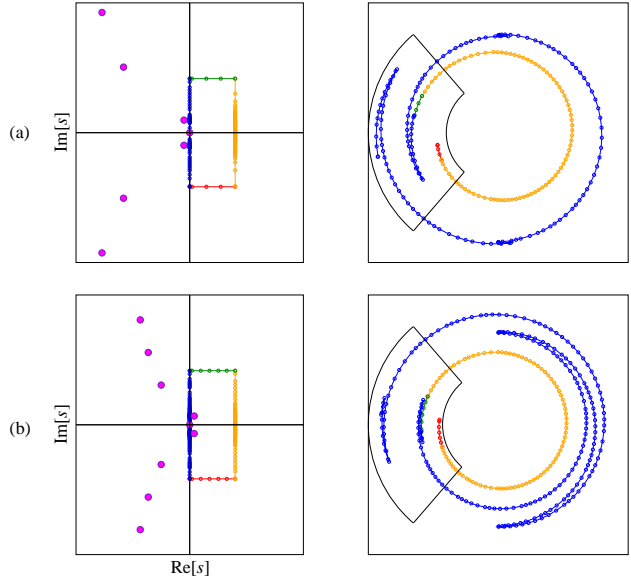}
\caption{(Color online) Numerical integration of Eq.~(\ref{Cauchy}) to identify
the presence of zeros in the cases of a system of two coupled nodes
($\tau_c$$=$$\pi/4$) for (a) $\tau$$=$$\pi/5$ and (b) $\tau$$=$$\pi/3$.
The left column shows the zeros and the points sampled along the
contour; the right column shows the argument of the characteristic
function (angular coordinate) at these steps (radial coordinate).}
\label{contour}
\end{figure}
Figure \ref{contour} shows two cases explored while finding the
critical delay. In Fig.~\ref{contour}(a), $\tau = \pi/5 < \tau_c$
and all real parts are non-positive so none fall within the contour.
Tracking the argument (right column) shows that the winding number
is correspondingly zero to verify that the delay is subcritical.
Alternatively, $\tau = \pi/3 > \pi_c$ in Fig.~\ref{contour}(b) and
there do indeed exist zeros with positive real parts that fall
within the contour. The argument winds around the origin twice,
signaling the presence of the first two zeros to cross the imaginary
axis, indicating instability.

\section{The Uniform Mode and the Width}
\label{appendix_UniformMode}

\subsection{Eigenmode Decomposition}

In synchronization and coordination problems, it is natural to define an observable such as the width, which measures fluctuations with respect to the global mean,
\begin{equation}
w^2(t) = \frac{1}{N} \sum_{i=1}^{N}[h_i(t)-\bar{h}(t)]^2 \;,
\end{equation}
where $\bar{h}(t)=\sum_{i=1}^{N}h_{i}(t)$.
In what follows, we show that the amplitude associated with the uniform mode of the normalized Laplacian automatically drops out from the width.
(In the case of unnormalized symmetric coupling, the expression for the width simplifies to the known form.)

For our problem with two types of time delays and locally normalized
couplings [Eq.~(\ref{diffEqNorm})], decomposition along the right
eigenvectors of $K^{-1}L$ facilitates diagonalization. While this
normalized Laplacian is a non-symmetric matrix, its eigenvalues are
all real and non-negative (with the smallest being zero,
$\lambda_0=0$). The corresponding (normalized) right eigenvector is
\begin{equation}
|e_{0}\rangle=N^{-1/2}(1,1,\ldots,1)^T \;.
\end{equation}
Note that since the normalized Laplacian is non-symmetric, the eigenvectors are not orthogonal, i.e., $\langle e_{l}|e_{k}\rangle\neq \delta_{lk}$.
To ease notational burden, in this subsection we use the bra-ket notation -- not to be confused with ensemble average over the noise.
In this notation, $\langle\cdot|$ is a row vector and $|\cdot\rangle$  is a column vector, e.g., $\langle e_{0}|=N^{-1/2}(1,1,\ldots,1)$.
Using this notation, the state vector is denoted by
\begin{equation}
|h(t)\rangle = (h_1(t),h_2(t),\ldots,h_N(t))^T \;,
\end{equation}
while the state vector relative to the mean is
\begin{eqnarray}
|h(t)-\bar{h}(t)\rangle & = & (h_1(t)-\bar{h}(t),h_2(t)-\bar{h}(t),\ldots,h_N(t)-\bar{h}(t))^T \nonumber \\
                        & = & (h_1(t),h_2(t),\ldots,h_N(t))^T -\bar{h}(t) (1,1,\ldots,1)^T \nonumber \\
                        & = & |h(t)\rangle - \bar{h}(t)\sqrt{N} |e_{0}\rangle
                          =   (1 -|e_0 \rangle\langle e_0|)|h(t)\rangle \;.
\end{eqnarray}
Employing the above formalism, the width can be written as
\begin{equation}
w^2(t) =
\frac{1}{N} \sum_{i=1}^{N}[h_i(t)-\bar{h}(t)]^2 =
\frac{1}{N} \langle h-\bar{h}| h-\bar{h} \rangle \;.
\end{equation}
Now we express the state vector as the linear combination of the eigenvectors of the underlying Laplacian,
\begin{equation}
|h(t)\rangle   \sum_{k=0}^{N-1} \tilde{h}_k(t) |e_{k}\rangle \;.
\end{equation}
Employing the above eigenmode decomposition, $\langle h-\bar{h}| h-\bar{h} \rangle$ can be written as
\begin{eqnarray}
\langle h-\bar{h}| h-\bar{h} \rangle & = & \langle h|(1 -|e_0 \rangle\langle e_0|)^2|h \rangle =
\langle h|(1 -|e_0 \rangle\langle e_0|)|h \rangle =
\sum_{k=0}^{N-1} \tilde{h}_k(t)\langle e_k|\left(1 -|e_0 \rangle\langle e_0|\right) \sum_{l=0}^{N-1} \tilde{h}_l(t)|e_l \rangle  \nonumber \\
& = & \sum_{k,l=0}^{N-1} \tilde{h}_k(t)\tilde{h}_l(t) \langle e_k|\left(1 -|e_0 \rangle\langle e_0|\right)|e_l \rangle =
\sum_{k,l=0}^{N-1} \tilde{h}_k(t)\tilde{h}_l(t) \left(\langle e_k|e_l \rangle - \langle e_k |e_0 \rangle\langle e_0|e_l \rangle \right) \nonumber \\
& = & \sum_{k,l\neq0} \tilde{h}_k(t)\tilde{h}_l(t) \left(\langle e_k|e_l \rangle - \langle e_k |e_0 \rangle\langle e_0|e_l \rangle \right) =
\sum_{k,l\neq0} \tilde{h}_k(t)\tilde{h}_l(t) \left( E_{kl} - E_{k0}E_{0l} \right) \;,
\label{hh}
\end{eqnarray}
where $E_{kl}\equiv\langle e_k |e_l \rangle$.
As can be seen explicitly from Eq.~(\ref{hh}), the terms where either $k$ or $l$ are zero drop out from the sum
(as $E_{00}$$=$$1$). It is also clear from Eq.~(\ref{hh}) that
$\langle h-\bar{h}| h-\bar{h} \rangle = \sum_{k\neq0} \tilde{h}_{k}^{2}(t)$ when the underlying coupling is symmetric
(and consequently the eigenvectors form an orthogonal set, $E_{kl}$$=$$\delta_{kl}$). Finally, the width can be written as
\begin{equation}
w^2(t) = \frac{1}{N} \sum_{i=1}^{N}[h_i(t)-\bar{h}(t)]^2 = \frac{1}{N} \langle h-\bar{h}| h-\bar{h} \rangle
= \frac{1}{N} \sum_{k,l\neq0} \tilde{h}_k(t)\tilde{h}_l(t) \left( E_{kl} - E_{k0}E_{0l} \right) \; .
\end{equation}
Note that the above result can be immediately applied to the case of symmetric coupling with no transmission delays [Eq.~(\ref{diffEqUniform})].
There, the eigenvectors of the corresponding Laplacian form an orthogonal set, and the above expression collapses to $w^2(t)=\frac{1}{N}\sum_{k=1}^{N-1} \tilde{h}_k^{2}(t)$ \cite{HuntPRL}.

\subsection{Ensemble Average over the Noise}

We now use the general form of the solution given in
Appendix~\ref{appendix_Fluctuations} [Eq.~(\ref{noise_delay_sol})] for the respective eigenmodes of normalized Laplacian coupling with two types of time delays [Eq.~(\ref{diffEqNormMode})], giving
\begin{equation}
\tilde{h}_{k}(t) = \int_0^t dt' \sum_\alpha\frac{e^{s_{k\alpha}(t - t')}}{g_{k}^{'}(s_{k\alpha})}\; \tilde{\eta}_{k}(t') \;,
\end{equation}
where $s_{k\alpha}$ is the $\alpha$th solution of the $k$th mode for the characteristic equation $g_{k}(s) = 0$
[Eq.~(\ref{char_eq_gk})].
After averaging over the noise, one obtains for the two-point function
\begin{equation}
\langle \tilde{h}_{k}(t) \tilde{h}_{l}(t) \rangle
= -2D \chi_{kl}\sum_{\alpha,\beta} \frac{( 1- e^{(s_{k\alpha} + s_{l\beta})t} )}{g_{k}^{'}(s_{k\alpha}) g_{l}^{'}(s_{l\beta})(s_{k\alpha} + s_{l\beta})} \;.
\end{equation}
In the stationary state, one must have ${\rm Re}(s_{k\alpha}) < 0$ for all $k$ and $\alpha$.
Thus, the stationary state width can be written as
\begin{equation}
\langle w^2(\infty) \rangle = \lim_{t\to\infty}
\frac{1}{N} \sum_{k,l\neq0} \langle \tilde{h}_k(t)\tilde{h}_l(t) \rangle \left( E_{kl} - E_{k0}E_{0l} \right) =
\frac{-2D}{N} \sum_{k,l\neq0} \sum_{\alpha,\beta}
\frac{\left( E_{kl} - E_{k0}E_{0l} \right)\chi_{kl}}{g_{k}^{'}(s_{k\alpha}) g_{l}^{'}(s_{l\beta})(s_{k\alpha} + s_{l\beta})} \;.
\label{hh_ss}
\end{equation}


\end{document}